\documentclass[twocolumn,tightenlines,floatfix,nofootinbib,showpacs,aps,epsf,amsfonts,amssymb,amsbsy]{revtex4}

\usepackage{graphics,bm}
\usepackage{epsfig}
\usepackage{graphicx}
\usepackage{rotating}

\newcommand{\beq}{\begin{equation}}
\newcommand{\eeq}{\end{equation}}
\newcommand{\bqa}{\begin{eqnarray}}
\newcommand{\eqa}{\end{eqnarray}}

\def\sumint{\hbox{$\sum$}\!\!\!\!\!\!\int}
\def\square{\vcenter{\vbox{\hrule height.4pt
          \hbox{\vrule width.4pt height4pt
          \kern4pt\vrule width.3pt}\hrule height.4pt}}}

\begin{document}

\author{Jens O. Andersen and Lars E. Leganger} 
\email{andersen@tf.phys.no}
\email{lars.leganger@ntnu.no}
\affiliation{Department of Physics, 
Norwegian Institute of Science and Technology, N-7491 Trondheim, Norway}
\title{Kaon condensation in the color-flavor-locked phase of
quark matter, the Goldstone theorem, and the 2PI Hartree approximation}

\date{\today}

\begin{abstract}
At very high densities, QCD is in the color-flavor locked phase, which
is a color-superconducting phase.
The diquark condensates break chiral symmetry in the same way as it
is broken in vacuum QCD and
gives rise to an octet of pseudo-Goldstone bosons and a superfluid mode.
The lightest of these are the charged and neutral kaons. 
For energies below the superconducting gap, 
the kaons are described by an 
$O(2)\times O(2)$-symmetric effective scalar field theory with 
chemical potentials. 
We use this effective
theory to study 
Bose-condensation of kaons and their 
properties as functions of the temperature and the chemical potentials.
We use the 2-particle irreducible effective action formalism in the
Hartree approximation. The renormalization of
the gap equations and the effective potential is studied in detail
and we show that the counterterms are independent of temperature and
chemical potentials.
We determine the
phase diagram and the medium-dependent quasiparticle masses.
It is shown that the Goldstone theorem is satisfied to a very good
approximation.
The effects of imposing electric charge neutrality is examined as well.

\end{abstract}
\pacs{11.15Bt, 04.25.Nx, 11.10Wx, 12.38Mh}

\maketitle


\section{Introduction}

There has been a large effort in recent years to map out the phase diagram
of QCD as a function of temperature and baryon chemical 
potential~\cite{frankie,frank,raja,dirk,misha,rob,mark,braunm}.
For example, much work has been done at high baryon density
and the understanding of this part of the phase diagram has improved
significantly as compared to one or two decades ago.
At sufficiently high density and low temperature, we know 
that QCD is in the color-flavor locked (CFL) 
phase~\cite{frankie,frank,raja,mark}. 
This state is a color
superconducting state since the quarks form Cooper pairs 
as electrons in an ordinary superconductor. 
The attraction between the quarks, which renders the Fermi surface 
unstable against the formation of Cooper pairs, is provided by one-gluon
exchange.

At asymptotically high densities, one can ignore the strange-quark mass
and quarks of all three colors and all three flavors participate in
a symmetric manner in the pairing.
The original symmetry group 
$SU(3)_{\rm c}\times SU(3)_L\times SU(3)_R \times U(1)_B$ 
is broken down
to $SU(3)_{c+L+R}$ which is a linear combination of the generators of the
original group. This linear combination locks rotation in color space with
rotations in flavor space and this has given the name to the phase.
In the CFL phase there is an octet of Goldstone modes which arises from 
the breakdown of chiral symmetry and a singlet arising from the breakdown
of the baryon-number group $U(1)_B$. The latter is a superfluid
mode since it is responsible for the superfluidity in the CFL phase.
This is analogous to the superfluidity encountered in Bose-Einstein condensed
phases
in condensed-matter systems.
Since the symmetry-breaking pattern is the same as in vacuum QCD, 
the low-energy properties 
of the CFL phase can be described in terms of an effective chiral Lagrangian
for the octet of (pseudo)-Goldstone modes and the superfluid
mode~\cite{gatto,sonny,kaon,kaon2,sjafer}. 
An important difference between chiral perturbation theory
in the vacuum and in the CFL phase is that the latter is at high density
and the Lagrangian is therefore coupled to chemical potentials via
the zeroth component of a "gauge field".

At asymptotically
high densities, all nine modes are exactly massless since one can
ignore the quark masses. 
At moderate densities, this is no longer the case.
The quark masses can not be neglected and chiral symmetry is explicitly
broken. This implies that only the superfluid mode is exactly
massless, while 
the other mesonic modes acquire masses. 
This is relevant for the interior of a neutron star.
In this case, the quark
chemical potential is of the order of 500 MeV, while the strange quark mass
is somewhere between the current quark mass of approximately 100 MeV
and the constituent quark mass of approximately 500 MeV~\cite{alfordus}.
The mass spectrum in the CFL is the opposite of vacuum QCD and
the lightest massive modes are expected to be the charged and neutral kaons
$K^+/K^-$ and $K^0/\bar{K}^0$. 
Furthermore if the chemical potential associated with
one of the bosons is larger than its vacuum mass, it will Bose condense.
Various aspects of condensation of kaons in the CFL phase
have been studied within Nambu-Jona-Lasinio 
models~\cite{buballanjl,forbes,ebertk,ebert+,ruggers,bubver,harmen,ebert3}.

Some of the properties of the kaons in the CFL phase have been studied
using effective scalar field theories. 
For example, it has been shown 
that the symmetry-breaking that accompanies Bose condensation of kaons
gives rise to unconventional Goldstone bosons if the original symmetry
group is $SU(2)\times U(1)$~\cite{shot,igor1,igor2}. This is the case when the 
chemical potential for the neutral kaons is the same as that of the charged
kaons. The Goldstone mode
has a quadratic dispersion relation for small values of the three-momentum
instead of the usual linear dispersion relation.
This unsual property of the Goldstone mode arises from the lack of Lorentz
invariance due to finite density.

In contrast to hadronic matter in heavy-ion collisions, bulk matter
in compact stars must (on average) be electrically neutral and so a
neutrality constraint must be imposed~\cite{coulomb,coulomb2}. 
Similarly, bulk matter must be color
neutral and if the system is in a color superconducting phase, 
one sometimes has to impose this constraint explicitly.
It is automatically satisfied if one uses the QCD Lagrangian, but
not so if one describes the system using NJL-type models.
If there are dynamical gauge fields present, the zeroth components of
$A^a_{\mu}$ develop nonzero expectation values, $\langle A^a_0\rangle\neq0$, 
that effectively act as chemical potentials.
In the NJL models there are no gauge fields present and 
the $SU(N_c)$ color symmetry is global~\cite{note,note2,blas,harmen}.
One must therefore impose these constraints explicitly.

One of the most popular approaches to the study of systems at finite
temperature and density is the 2-particle irreducible (2PI)
effective action formalism
developed by Cornwall, Jackiw and Tomboulis~\cite{cjt} in the context
of relativistic field theories. See e.g. 
Refs.~\cite{amelino,petro,kirk,BP-01,jmqed,bergesscalar,borsa,joalargen}.
for various applications.
The 2PI functional depends on 
background fields $\phi_0$ and the exact propagator $D$.
In practical calculations one must
truncate the exact generating 2PI functional according to some
approximation scheme. The $1/N$-expansion and the loop expansion are
two such systematic schemes. These approximation schemes are nonperturbative
in the sense that they sum up loop diagrams from all orders in the loop
expansion. The renormalization of such approximations is therefore 
a nontrivial issues as standard theorems from perturbation theory do not 
apply. There has been significant progress regarding the renormalization 
of these approximations 
starting with the papers by van Hees and Knoll~\cite{hees}.
They showed that the equation for the
two-point function in scalar $\phi^4$-theory
can be renormalized by introducing a finite number
of counterterms. A more systematic study of the renormalization issues
in scalar $\phi^4$-theory was presented in Ref.~\cite{bir3}. 
They formulated an iterative renormalization procedure that determines
the counterterms needed to eliminate the (sub)divergences. These counterterms
were shown to be independent of temperature.
In Ref.~\cite{fejos} the authors have developed a ``direct" renormalization
procedure which is equivalent to the iterative procedure in~\cite{bir3}.
Later it has been shown~\cite{berges2} how to 
fix all the counterterms needed to calculate the proper vertices which
are encoded in the effective action. These counterterms are local and 
they are independent of temperature and chemical potential.

The 2PI effective action formalism for scalar fields has recently
been used to study the thermodynamics
of pions and  kaons and their condensation.
In Ref.~\cite{joalargen}, the
quasi-particle masses and the phase diagram is studied to leading order
in the $1/N$ expansion of $O(N)$-symmetric scalar field theories.
In Ref.~\cite{alfordus}, the authors 
applied the 2PI effective action formalism 
in the Hartree approximation to an effective $O(2)\times O(2)$-symmetric
scalar field and calculated the phase diagram
and the critical temperature for Bose-condensation of kaons.
The effects of imposing
electric charge neutrality were also investigated.
The scalar theory for the kaons were derived
from the effective chiral 
Lagrangian, where the parameters depend on the baryon chemical potential.
Renormalization issues were not addressed.
In the present paper, we reconsider the problem of kaon condensation
from a somewhat different angle, and consider in some detail   
the renormalization of the theory. We also show that the violation
of Goldstone's theorem is negligible.

The paper is organized as follows. In Sec.~II, we discuss the 2PI 
Hartree approximation for a $O(N)$-symmetric Bose gas. 
In Sec.~III, we briefly discuss $O(2)\times O(2)$-symmetric 
models, which are relevant 
for kaon condensation in the CFL phase of dense quark matter. We determine
the quasi-particle masses as well as the phase diagram. Finally, we
study the effects of imposing electric charge neutrality. In Sec.~IV
we summarize and conclude. In appendix A, we discuss the renormaliztion
of the gap equations and effective potential in detail.

\section{$O(2N)$-symmetric Bose gas}
The Euclidean
Lagrangian for a Bose gas with $N$ species of massive charged scalars is
\bqa
{\cal L}&=&
(\partial_{\mu}\Phi_i^{\dagger})(\partial_{\mu}\Phi_i)
+m^2\Phi^{\dagger}_i\Phi_i
+{\lambda\over2N}\left(\Phi^{\dagger}_i\Phi_i\right)^2
\;,
\label{lag}
\eqa
where $i=1,2,...,N$ and 
$\Phi_i=(\phi_{2i-1}+i\phi_{2i})/\sqrt{2}$ is a complex field.
The theory described by Eq.~(\ref{lag})
has $(2N-1)N$ conserved charges which equals the number
of generators of the group $O(2N)$. 

A gas of $N$ species of  
bosons can be characterized by the expectation values
of the different conserved charges in addition to the temperature.
For each conserved charge $Q_i$, one may introduce a nonzero
chemical potential $\mu_i$. However, it is possible to specify the 
expectation values of different charges only if they commute.
The maximum number of commuting charges is $N$~\cite{arthur} 
and these can 
be chosen as 
\bqa
Q_{i}&=&\int\;d^3x\,j_{i}^0\;,
\eqa
where the the current densities $j_{i}^{\mu}$ are
\bqa
j_i^{\mu}&=&\phi_{2i}\partial^{\mu}\phi_{2i-1}
-\phi_{2i-1}\partial^{\mu}\phi_{2i}
\eqa
The incorporation of a conserved charge $Q_i$ is done by making the substitution
\bqa
\partial_0\Phi_i\rightarrow\left(\partial_0-\mu_i\right)\Phi_i\;\\ 
\partial_0\Phi_i^{\dagger}\rightarrow\left(\partial_0+\mu_i\right)
\Phi_i^{\dagger}\;.
\eqa
in the Lagrangian~(\ref{lag}). Note that the chemical potential acts
as the zeroth component of a gauge field.

From the path-integral representation of the thermodynamic potential
$\Omega$
\bqa
e^{-\beta V\Omega}&=&\int{\cal D}\Phi^*_i{\cal D}\Phi_i 
e^{-\int_0^{\beta}d\tau\int d^3x\cal L}\;,
\eqa
the expression for the charge density can be written as
\bqa
Q_i&=&-{\partial{\Omega}\over\partial\mu_i}\,.
\eqa
If we introduce $k$ chemical potentials, the full symmetry group is
broken down to $[O(2)]^k\times O(2N-2k)$.
If $m^2<0$, the $O(2N)$ symmetry is spontaneously broken down to
$O(2N-1)$. Even if $m^2>0$, the symmetry may be broken if 
one of the chemical potentials, $\mu_i$, 
is larger than a critical chemical potential $\mu_c=m$.
In that case, the $\langle0|\phi_{2i-1}|0\rangle\neq0$ and the corresponding
$O(2)$ symmetry is broken.
In the following, we consider the simplest example of a single chemical
potential $\mu=\mu_i$ for the complex field $\Phi_1$.
We first introduce a nonzero vacuum expectation value $\phi_0$ for the
field $\Phi_1$
in order to allow for a charged condensate. 
Using the $O(2)$-symmetry, we can always choose $\phi_0$ real and so we
can write
\bqa
\Phi_1&=&{1\over\sqrt{2}}\left(\phi_0+\phi_1+i\phi_2\right)\;,
\eqa
where $\phi_1$ and $\phi_2$ are quantum fluctuating fields.
\begin{widetext}
The inverse tree-level propagator then reads
\bqa
D_0^{-1}(\omega_n,p)=
\left(\begin{array}{ccccc}
\omega_n^2+p^2+m_1^2-\mu^2&-2\mu\omega_n&0&0&...
\\
2\mu\omega_n&\omega_n^2+p^2+m_2^2-\mu^2&0&0&...
\\0&0&\omega_n^2+p^2+m_3^2&0&...
\\
0&0&0&\omega_n^2+p^2+m_3^2&... \\
...&...&...&...&...\\
\end{array}\right)\;,
\label{eq:D0}
\eqa
\end{widetext}
where $\omega_n=2\pi nT$ are the Matsubara frequencies and 
the tree-level masses are
\bqa
m_1^2&=&m^2+{3\lambda\over2N}\phi_0^2\;,\\
m_2^2&=&m^2+{\lambda\over2N}\phi_0^2\;, \\
m_3^2&=&m^2+{\lambda\over2N}\phi_0^2\;.
\eqa
The tree-level dispersion relation is found by analytic continuation to
Minkowski space, $\omega_n\rightarrow i\omega$, and then solving the
equation ${\rm Det}D_0(\omega,p)=0$. This yields
\begin{widetext}
\bqa
\label{mpm22}
\omega_{1,2}(p)&=&\sqrt{p^2+{1\over2}(m_1^2+m_2^2)+\mu^2\pm 
\sqrt{4\mu^2\left[p^2+\frac{1}{2}(m_1^2+m_2^2)\right] + 
\frac{1}{4}\left(m_1^2-m_2^2\right)^2}} \;, \\
\omega_3(p)&=&\sqrt{p^2+m_3^2}\;.
\label{mpm3}
\eqa
\end{widetext}
The classical effective potential is
\bqa
V&=&{1\over2}\left(m^2-\mu^2\right)\phi_0^2
+{\lambda\over8N}\phi_0^4
\;.
\eqa
The minimum of the classical potential $V$ is given by
$(m^2-\mu^2)+\lambda/2N\phi_0^2=0$ and at the minimum, 
we have 
$m_1^2=3\mu^2-2m^2$ and $m_2^2=\mu^2$.
The dispersion relation $\omega_{1,2}(p)$ then reduces to
\bqa\nonumber
\omega_{1,2}(p)&=&
\sqrt{p^2+3\mu^2-m^2\pm\sqrt{(3\mu^2-m^2)^2+4\mu^2 p^2}}\;,
\\ &&
\label{pmmode}
\eqa
From this equation, we note that $\omega_2(p)$ is a massless mode. Expanding
around zero momentum $p$, we find
\bqa
\omega_2(p)&=&\sqrt{\mu^2-m^2\over3\mu^2-m^2}\,p
+{\cal O}(p^2)\;.
\label{plusmode}
\eqa
The mode is linear in the momentum for small momenta and is thus a 
conventional Goldstone mode. This is in agreement with the fact that
the $O(2)$-symmetry has been broken due to the condensate $\phi_0$.
\subsection {Effective Action and Gap Equations}
\begin{figure}[htb]
\begin{center}
\includegraphics[width=5.7cm]{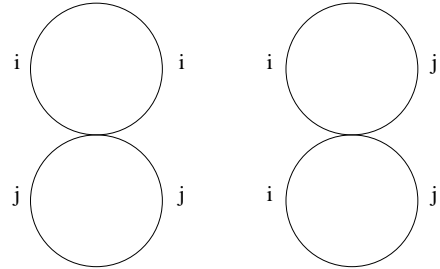}
\caption{Vacuum diagram contributing 
to the effective potential.in the Hartree approximation.
Left diagrams are of order $N$, while right diagrams are of order one.} 
\label{bubble22}
\end{center}
\end{figure}

The 2PI effective action can be written as
\bqa\nonumber
\Omega[\phi_0,D]&=&
{1\over2}\left(m^2-\mu^2\right)\phi_0^2
+{\lambda\over8N}\phi_0^4 
+{1\over2}{\rm Tr}\ln D^{-1}
\\&&
+{1\over2}{\rm Tr}D_0^{-1}D
+\Phi[D]\;,
\label{2piea}
\eqa
where $D$ is the exact propagator and $\Phi[D]$ is the sum of all two-particle
irreducible vacuum diagrams. The trace is over field indices as well as
over space-time. 
In the Hartree approximation one includes the double bubble diagrams
shown in Fig.~\ref{bubble22}. If we denote by $D_{ij}$ the components of the
propagator $D$, the functional $\Phi[D]$ can be written as 
\bqa
\Phi[D]&=&{\lambda\over8N}
F_{ijkl}\sumint_PD_{ij}\sumint_QD_{kl}\;,
\eqa
where 
$F_{ijkl}=(\delta_{ij}\delta_{kl}+\delta_{ik}\delta_{jl}+\delta_{il}\delta_{jk})$
is the sum of the three rank-four invariants of $O(N)$.
The sum-integral above is defined by
\bqa
\sumint_Q&\equiv&\left({e^{\gamma_E}\Lambda^2\over4\pi}\right)^{\epsilon}
T\sum_{q_0=2\pi nT}\int{d^dq\over(2\pi)^d}\;,
\eqa 
where $Q=(q_0,{\bf q})$,
$d=3-2\epsilon$ and $\Lambda$ is the renormalization scale associated
with dimensional regularization. The sum is over Matsubara frequencies.
The integral over three-momentum $q$
is calculated with dimensional regularization.
We also introduce the compact notation for these integrals
\bqa
\int_q&\equiv&\int{d^dq\over(2\pi)^d}\;.
\eqa
Generally, in the vacuum, the terms in the 
2PI effective action~(\ref{2piea}) can be classified according
to which order in the $1/N$-expansion they contribute and can be 
expressed in terms of $O(N)$ invariants such as ${\rm Tr}(D^n)$
and ${\rm Tr}(\phi_0^2D^n)$~\cite{gertie}. 
The first and second term 
in Eq.~(\ref{2piea}) 
are proportional to ${\rm Tr}[\phi^2_0]$
and $[{\rm Tr}(\phi^2_0)]^2$, respectively. Each trace gives a factor of $N$
and so they both scale as $N$. The third term is the trace of the full
propagator and therefore scales as $N$.
The third term can be decomposed into a sum of the terms
${\rm Tr}[D]$, ${\rm Tr}[\phi^2_0D]$, ${\rm Tr}[\phi^2_0]{\rm Tr}[D]$, and
${\rm Tr}[\phi^2_0D]$.
Finally, let us consider $\Phi[D]$.
Performing the sums involving $F_{ijkl}$, 
$\Phi[D]$ can be expressed in terms of the
invariants ${\rm Tr}(D^2)$ and $[{\rm Tr}(D)]^2$:
\bqa
\Phi[D]&=&{\lambda\over8N}\bigg[
[{\rm Tr}\left(D\right)]^2+2{\rm Tr}\left(D^2\right)
\bigg]\;.
\eqa
The first term involves two traces each giving a factor of $N$. There is
a factor of $1/N$ coming from the vertex and so this term goes like $N$.
Similarly, the second term involves one trace and one vertex and 
it contributes at order one. We therefore conclude that the Hartree
approximation is not a systematic approximation. It is not systematic
in powers of $1/N$ as we have seen and it is not systematic in number
of loops since we have not included the setting-sun diagrams that arise
in the broken phase.

\begin{figure}[htb]
\includegraphics[width=5.cm]{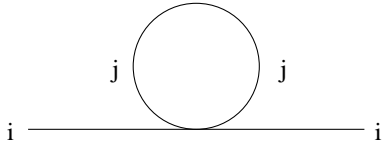}
\caption{
Feynman diagram contributing to the self-energy $\Pi(P)$
in the Hartree approximation.}
\label{selfy}
\end{figure}

At the stationary points, the 2PI effective action satisfies the gap equations
\bqa
{\delta\Omega[\phi_0,D]\over\delta D}&=&0\;,
\label{latter}
\\
\label{former}
{\delta\Omega[\phi_0,D]\over\delta\phi_0}&=&0\;.
\eqa
The gap equation~(\ref{latter}) can be rewritten as
\bqa
D^{-1} = D_0^{-1} + 2{\delta\Phi[D]\over\delta D}\;.
\label{eq:D}
\eqa
Using the fact that $D^{-1}-D_0^{-1}=\Pi(P)$, where $\Pi(P)$ is the
self-energy, we obtain
\bqa
\Pi(P)&&= 2{\delta\Phi[D]\over\delta D}\;.
\label{pip}
\eqa
Note that the self-energy in Eq.~(\ref{pip}) is a matrix.
The self-energy is obtained by cutting a line in the vacuum graphs.
A generic loop diagram that contributes in the Hartree approximation is shown in
Fig.~\ref{selfy}.

In the Hartree approximation, the self-energies are momentum independent. 
Moreover, it can be shown~\cite{alfordus} that only the diagonal
elements of the self-energy matrix $\Pi_{ij}$ are nonzero.
If we denote these elements by $\Pi_i$, the dressed inverse propagator 
is given by Eq.~(\ref{eq:D0}) with the tree level masses $m_1,\:m_2,\;m_3$ 
replaced by the effective medium-dependent masses
$M_1,\;M_2,\;M_3$, where
\bqa
M_i^2&=&m_i^2+\Pi_{i}\;.
\label{dressm}
\eqa 
The full propagator then reads
\begin{widetext}
\bqa
D(\omega_n,p)
&=&
\left(\begin{array}{ccccc}
{\omega_n^2+p^2+M_1^2-\mu^2
\over(\omega_n^2+\tilde{\omega}_1^2)(\omega_n^2+\tilde{\omega}_2^2)}
&{2\mu\omega_n
\over(\omega_n^2+\tilde{\omega}_1^2)(\omega_n^2+\tilde{\omega}_2^2)}
&0&0&...
\\
{-2\mu\omega_n\over(\omega_n^2+\tilde{\omega}_1^2)(\omega_n^2+\tilde{\omega}_2^2)}
&{\omega_n^2+p^2+M_2^2-\mu^2
\over(\omega_n^2+\tilde{\omega}_1^2)(\omega_n^2+\tilde{\omega}_2^2)}
&0&0&...
\\0&0&{1\over\omega_n^2+\tilde{\omega}_3^2}&0&...
\\
0&0&0&{1\over\omega_n^2+\tilde{\omega}_3^2}&... \\
...&...&...&...&...
\label{fulld}
\end{array}\right)\;,
\eqa
where $\tilde{\omega}_{1,2,3}(p)$ are 
obtained from Eqs.~(\ref{mpm22}) and~(\ref{mpm3}) by the replacement
$m_i\rightarrow M_i$.
The functional $\Phi[D]$ can then be written out explcitly as
\bqa\nonumber
\Phi[D]&=&{\lambda\over8N}
\left[
3\sumint_QD_{11}\sumint_KD_{11}+3\sumint_QD_{22}\sumint_KD_{22}
+2\sumint_QD_{11}\sumint_KD_{22}
+4(N-1)\sumint_QD_{11}\sumint_KD_{33}
\right.\\ &&\left.
+4(N-1)\sumint_QD_{22}\sumint_KD_{33}
+4N(N-1)\sumint_QD_{33}\sumint_KD_{33}
\right]\;.
\label{phiex}
\eqa
Notice that the terms involving off-diagonal elements of the full propagator
$D$ are absent. This is due to the fact the diagram vanishes
upon summation over the Matsubara frequencies. This follows
immediately from Eq.~(\ref{fulld}).

The gap equations for the dressed masses now follow from 
Eqs.~(\ref{eq:D}),~(\ref{pip}),~(\ref{dressm}), and~(\ref{phiex}).
These equations contain ultraviolet divergences and require renormalization. 
Renormalization is discussed in Appendix A and 
the diagrammatic interpretation of the iterative renormalization procedure
is shown in Fig.~\ref{hartree22}.

\begin{figure}[htb]
\includegraphics[width=11.0cm]{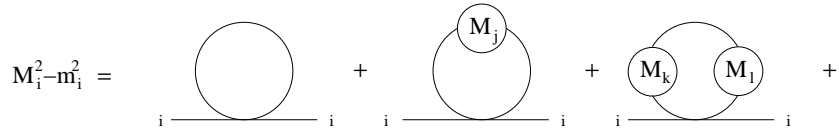}
\vspace{-1.5cm}
\caption{Diagrammatic interpretation of the gap equations 
for $m_i$ in the 2PI
Hartree approximation.}
\label{hartree22}
\end{figure}

After renormalization, we obtain
\bqa
\label{m1dress1}
M^2_1 - m^2_1 &=& {\lambda\over2N}\left[
3(J^c_1 + J^T_1) + (J^c_2 + J^T_2) + (2N-2)(J^c_3 + J^T_3)\right]
\;,\\
\label{m2dress2}
M^2_2-m^2_2 &=& {\lambda\over2N}\left[
(J^c_1 + J^T_1) + 3(J^c_2 + J^T_2) + (2N-2)(J^c_3 + J^T_3)\right]
\;,\\
M^2_3-m^2_3 &=& {\lambda\over2N}\left[
(J^c_1 + J^T_1) + (J^c_2 + J^T_2) + 2N(J^c_3 + J^T_3)\right]
\;,
\label{m3dress3}
\eqa
where the integrals $J_n^c$ and $J_n^T$ are defined in Appendix A.

We next consider the gap equation~(\ref{former}). 
The expression for it follows from~(\ref{latter}) and~(\ref{phiex}).
After renormalization, 
we obtain
\bqa
0 &=& \phi_0\left[m^2-\mu^2+{\lambda\over
2N}\phi_0^2+{\lambda\over2N}\left[3(J^c_1+J^T_1) + (J^c_2+J^T_2) +
(2N-2)(J^c_3+J^T_3)\right]\right]\;.
\label{phidress}
\eqa
\end{widetext}
Using the gap equation~(\ref{m1dress1}) to eliminate the integrals
$J_n^c$ and $J_n^T$,
we can rewrite Eq.~(\ref{phidress}) in a very simple way
\bqa
0&=&\phi_0\left[M_1^2-\mu^2-{\lambda\over N}\phi_0^2
\right]\;.
\eqa
The difference between this equation and the corresponding
equation at the tree-level, is the replacements $m_1^2\rightarrow M_1^2$.
Similarly, combining the gap equations~(\ref{m2dress2}) and~(\ref{phidress}), 
we obtain
\bqa
M_2^2&=&\mu^2+
{\lambda\over N}
\left[
(J_2^c+J_2^T)-(J_1^c+J_1^T)
\right]\;.
\label{signal}
\eqa
Comparing this equation with the tree-level result $m_2^2=\mu^2$, 
Eq.~(\ref{signal}), we see that the Goldstone theorem is not respected.
It is well known that the 2PI Hartree approximation 
violates Goldstone's theorem and 
in the present case this means that there is no massless mode associated with
the breaking of the $O(2)$-symmetry due to the condensate $\phi_0$.
Writing $M_2^2=\mu^2+\delta$, where 
$\delta={\lambda\over N}
\left[
(J_2^c+J_2^T)-(J_1^c+J_1^T)
\right]$, a calculation analogous to the one leading to Eq.~(\ref{plusmode}),
shows that the mass gap of the Goldstone mode is given by
\bqa
\Delta M^2_{\rm GB}&=&{M_1^2-m^2\over3\mu^2+M_1^2}\delta
\;.
\eqa
In the 2PI $1/N$-expansion, the Goldstone theorem is satisfied order by order.
In the large-$N$ limit, it is clear from Eq.~(\ref{signal}) 
that the Goldstone theorem is respected
since the last term vanishes.

The effective potential~(\ref{2piea}) 
follows from Eqs.~(\ref{eq:D0}), ~(\ref{fulld}) and~(\ref{phiex})
It can be renormalized in the same manner as  
the gap equations and details can be found in Appendix A. After 
renormalization, we obtain
\begin{widetext}
\bqa\nonumber
\Omega&=&{1\over2}(m^2-\mu^2)\phi_0^2+{\lambda\over8N}\phi_0^4
+ \frac{1}{2}({\cal J}^c_1 + {\cal J}^T_1)
+ \frac{1}{2}({\cal J}^c_2 + {\cal J}^T_2)
+ (N-1)({\cal J}^c_3 + {\cal J}^T_3)\\ & &
-\frac{1}{2}(M_1^2-m_1^2)({J}^c_1 + { J}^T_1)
-\frac{1}{2}(M_2^2-m_2^2)({ J}^c_2 + { J}^T_2)
-(N-1)(M_3^2-m_3^2)({J}^c_3 + { J}^T_3)\nonumber\\
\nonumber & & + {3\lambda\over 8N}({J}^c_1+J^T_1)^2
+ {3\lambda\over 8N}({J}^c_2+J^T_2)^2
+{\lambda\over 2}(N-1)({J}^c_3+J^T_3)^2 
+ {\lambda\over 4N} 
({J}^c_1+J^T_1)({J}^c_2+J^T_2)
\\  & & 
+{\lambda\over 2N}(N-1) 
({J}^c_1+J^T_1)({J}^c_3+J^T_3) 
+ {\lambda\over 2N}(N-1) 
(J^c_2+J^T_2)(J^c_3+J^T_3)\;,
\label{onomega}
\eqa
\end{widetext}
where ${\cal J}_n^c$ and ${\cal J}_n^T$ are defined in Appendix A.

\section{Kaons in the  CFL phase}

\subsection{Effective theories}
In the CFL phase of dense QCD, the original symmetry 
$SU(3)_c\times SU(3)_L\times SU(3)_R\times U(1)_B$ is broken down to 
$SU(3)_{c+L+R}$. The diquark condensate $\langle\psi\psi\rangle$
breaks chiral symmetry in exactly
the same manner as in vacuum QCD and so the effective Lagrangian for the
Goldstone modes have the same structure as in chiral perturbation theory.
Notice, however, that the mesons are composed of four quark fields of the
form $\bar{\psi}\bar{\psi}\psi\psi$ instead of the conventional 
$\bar{\psi}\psi$. Nevertheless, one finds the same quantum numbers.
Another difference is that Lorentz invariance is broken due to the
presence of the chemical potentials and so 
the Lagrangian in invariant only under rotations in three dimensions.
The chiral effective Lagrangian is given by~\cite{sonny}
\bqa\nonumber
{\cal L}&=&{1\over4}f_{\pi}^2{\rm Tr}
\left[
\left(\partial_0\Sigma+i[A,\Sigma]\right)
\left(\partial_0\Sigma-i[A,\Sigma]^{\dagger}\right)
\right. \\ &&\nonumber\left.
-v_{\pi}^2(\partial_i\Sigma)(\partial_i\Sigma^{\dagger})
\right] 
\\ &&
+{1\over2}af_{\pi}^2\det M{\rm Tr}[M^{-1}(\Sigma+\Sigma^{\dagger})]\;+\cdots,
\eqa
where $f_{\pi}$, $v_{\pi}$, and $a$ are constants.
The meson field $\Sigma$ is given by
\bqa
\Sigma&=&e^{i\lambda^a\phi^a/f_{\pi}}\;,
\eqa
where $\lambda^a$ are the Gell-Mann matrices and $\phi^a$ describe
the octet of Goldstone bosons.
The matrix $A$ acts as the zeroth component of a gauge field and is given by 
\bqa
A&=&\mu_QQ-{M^2\over2\mu}\;,
\eqa
where $\mu_Q$ is the chemical potential for electric charge, $\mu$
is the baryon chemical potential, $Q={\rm diag}(2/3,-1/3,-1/3)$, and
$M={\rm diag}(m_u,m_d,m_s)$.

At asymptotically high densities, one can use perturbative QCD calculations
to determine the parameters $f_{\pi}$, $v_{\pi}$, and $a$ by 
matching~\cite{kaon2,sonny,sjafer}. 
\bqa
\label{fasym}
f_{\pi}^2&=&{12-8\ln2\over18\pi^2}\mu^2\;, \\
v_{\pi}^2&=&{1\over3}\;,\\
a&=&{3\Delta^2\over\pi^2 f_{\pi}^2}\;,
\label{aasym}
\eqa
where $\Delta$ is the superconducting gap.
Note that $v_{\pi}=1/\sqrt{3}$ is the standard result for the 
speed of sound in a dense medium. 
In the vacuum, Lorentz invariance enforces the value $v_{\pi}=1$.
It is important to stress that the values
of these parameters given by Eqs.~(\ref{fasym})--(\ref{aasym}) are valid
only at asympotically high densites. For moderate densities, which are
relevant for compact stars, one does not know the values of
these parameters as they cannot be determined by 
matching.

After expanding to fourth order in the meson fields, 
Alford, Braby, and Schmitt 
obtain the following effective Lagrangian for the kaons:
\begin{widetext}
\bqa\nonumber
{\cal L}&=&
\left[(\partial_{0}+\mu_1)\Phi_1^{\dagger}\right]\left[(\partial_{0}-\mu_1)
\Phi_1\right]
+(\partial_{i}\Phi_1^{\dagger})(\partial_{i}\Phi_1) 
+ \left[(\partial_{0}+\mu_2)\Phi_2^{\dagger}\right]
\left[(\partial_{0}-\mu_2)\Phi_2\right]
+{1\over2} 
m_1^2
\Phi_1^2
\\ &&
+{1\over2}  
m_2^2 
\Phi_2^2
+{\beta_1\over4}\Phi_1^4+{\beta_2\over4}\Phi_2^4
+{\alpha\over2}\Phi_1^2\Phi_2^2
\;,
\label{elalford}
\eqa
\end{widetext}
where the complex doublet are $(K^+,K^0)=(\Phi_1,\Phi_2)$ and 
the parameters are given by
\bqa
\label{pam1}
m_1^2&=&am_d(m_s+m_u) \\
m_2^2&=&am_u(m_s+m_d) \\
\mu_1&=&\mu_Q+{m_s^2-m_u^2\over2\mu} \\ 
\label{pam4}
\mu_2&=&{m_s^2-m_d^2\over2\mu} \\ 
\label{pam5}
\beta_i&=&{1\over6f^2_{\pi}}\left(4\mu_i^2-m_i^2\right)\;,\\
\alpha&=&{1\over2}\left(\beta_1+\beta_2\right)
-\left({\mu_1-\mu_2\over2f_{\pi}}\right)^2\;.
\label{paml}
\eqa
The dimensionless parameters $\alpha$ and $\beta_i$ are quartic couplings
of the effective theory. Note in particular that they depend on the chemical
potentials $\mu_i$. The effective theory described by the 
Lagrangian~(\ref{elalford}) is invariant under the group
$O(2)\times O(2)$. 

There is a technical complication arising from using the 
Lagrangian Eq.~(\ref{elalford}). The problem is that the effective 
couplings depend on the chemical potentials. 
This implies that the counterterms also depend on the chemical potentials
i.e. parameters that describe a dense medium.
Renormalizing a theory based on the Lagrangian~(\ref{elalford})
therefore depends on the medium, which one may object to.
We therefore take a somewhat different approach by using an effective
Lagrangian with mass parameters and couplings that are independent of
the chemical potentials. For simplicity, we set $v_{\pi}=1$, but it is not
difficult to scale loop momenta in our equations to take into account 
values of $v_{\pi}$ that differ from the its vacuum value.
The kaons are 
written as a complex doublet~\footnote{Note that the identification
in Ref.~\cite{alfordus} is $(K^+,K^0)=(\Phi_1,\Phi_2)$.}, 
$(K^0,K^+)=(\Phi_1,\Phi_2)$. 
The Euclidean Lagrangian with an $O(2)\times O(2)$ symmetry is given by
\bqa\nonumber
{\cal L}&=&
\left[(\partial_{0}+\mu_0)\Phi_1^{\dagger}\right]\left[(\partial_{0}-\mu_0)\Phi_1\right]
+(\partial_{i}\Phi_1^{\dagger})(\partial_{i}\Phi_1) \\ \nonumber & & 
+ \left[(\partial_{0}+\mu_+)\Phi_2^{\dagger}\right]\left[(\partial_{0}-\mu_+)\Phi_2\right]
+(\partial_{i}\Phi_2^{\dagger})(\partial_{i}\Phi_2) \\ \nonumber & &
+m_0^2\Phi^{\dagger}_1\Phi_1+m_+^2\Phi^{\dagger}_2\Phi_2
+{\lambda_0\over2}\left(\Phi^{\dagger}_1\Phi_1\right)^2
\\ & &+{\lambda_+\over2}\left(\Phi^{\dagger}_2\Phi_2\right)^2 
+{\lambda_H}\left(\Phi^{\dagger}_1\Phi_1\right)\left(\Phi^{\dagger}_2\Phi_2\right)
\;.
\label{lag2x2}
\eqa
The chemical potentials
$\mu_0$ and $\mu_+$ associated with the two conserved charges for
each complex field $\Phi_i$. They are related to the 
quark chemical potentials $\mu_u$, $\mu_d$, and $\mu_s$ by
\bqa
\mu_0&=&\mu_d-\mu_s \;, \\
\mu_+&=&\mu_u-\mu_s \;. 
\eqa
We therefore have $\mu_+-\mu_0=\mu_u-\mu_d$. In weak 
equilibrium, the processes $d+\nu\leftrightarrow u+e^-$ go with the same rate in
both directions. If we assume that the neutrinos leave the system, their
chemical potential is $\mu_{\nu}=0$. This implies that 
$\mu_d=\mu_u-\mu_Q$, where $\mu_Q$ is the electric charge chemical potential.
In other words, $\mu_Q=\mu_+-\mu_0$.
In the absence of the chemical potentials and with $m_0=m_+$, and
$\lambda_0=\lambda_+=\lambda_H$, the Lagrangian~(\ref{lag2x2})
has an extended $SO(4)\sim SU(2)_L\times SU(2)_R$ symmetry.
If we add chemical potentials such that 
$\mu_0=\mu_+$, this symmetry is broken down
to $SU(2)\times U(1)$, where the
chemical potential is for the $U(1)$ charge. 
A condensate would then break this symmetry down to
$U(1)$ implying the existence of massless modes.
Naively, one would perhaps expect three Goldstone modes as there are three
broken generators. However, one of the massless mode is quadratic in the
momentum $p$ for small $p$ and the Nielsen-Chadha theorem implies that
such a mode be counted twice~\cite{holger,igor2,brauner}. 
This is consistent with the fact that there
are only two massless modes.
\subsection{Effective potential and gap equations}
In order to allow for a condensate of neutral kaons, we introduce
an expectation value $\phi_0$ for $\Phi_1$ and write
\bqa
\Phi_1&=&
{1\over\sqrt{2}}\left(\phi_0+\phi_1+i\phi_2\right)
\;,
\eqa
where $\phi_1$ and $\phi_2$ are quantum fluctuting fields.
\begin{widetext}
The inverse tree-level propagator can be written as a block-diagonal 
$4\times4$ matrix:
\bqa
D_0^{-1}(\omega_n,p)=
\left(\begin{array}{cccc}
\omega_n^2+p^2+m_1^2 - \mu_0^2&-2\mu_0\omega_n&0&0
\\
2\mu_0\omega_n&\omega_n^2+p^2+m_2^2 - \mu_0^2&0&0
\\0&0&\omega_n^2+p^2+m_3^2 - \mu_+^2&-2\mu_+\omega_n
\\
0&0&2\mu_+\omega_n&\omega_n^2+p^2+m_3^2 - \mu_+^2 \\
\end{array}\right)\;,
\eqa
\end{widetext}
where the tree-level masses are
\bqa
m_1^2&=&m_0^2+{3\lambda_0\over 2}\phi_0^2\;,\\
m_2^2&=&m_0^2+{\lambda_0\over 2}\phi_0^2\;, \\
m_3^2&=&m_+^2+{\lambda_H\over 2}\phi_0^2\;.
\eqa
Note that in the remainder of this section, the mass parameters 
$m^2_0$ and $m^2_+$ are 
positive. The classical potential is
\bqa
V={1\over2}\left(m_0^2-\mu_0^2\right)\phi_0^2+{\lambda_0\over8}\phi_0^4\;.
\eqa
\begin{widetext}
The dispersion relations are
\bqa
\omega_{1,2}(p)&=&\sqrt{p^2+{1\over2}(m_1^2+m_2^2)+\mu_0^2\pm 
\sqrt{4\mu_0^2\left[p^2+\frac{1}{2}(m_1^2+m_2^2)\right] + 
\frac{1}{4}\left(m_1^2-m_2^2\right)^2}} \;, \\
\label{mpm}
\omega_{3,4}(p)&=&\sqrt{p^2+m_3^2}\pm\mu_+
\;.
\eqa

The 2PI effective action is given by
\bqa
\Omega[\phi_0,D]&=&
{1\over2}
\left(m^2-\mu_0^2\right)\phi_0^2
+{\lambda\over8N}\phi_0^4 
+{1\over2}{\rm Tr}\ln D^{-1}
+{1\over2}{\rm Tr}D_0^{-1}D
+\Phi[D]\;,
\eqa
where $\Phi[D]$ contains all 2PI vacuum diagrams.
In the Hartree approximation, we include all double-bubble diagrams
which can be written in terms of $O(2)\times O(2)$ invariants.
If we denote by $D_a$ and $D_b$ the two $2\times2$ submatrices of the
propagator $D$, we can write~\cite{fejos}
\bqa
\Phi[D]&=&{\lambda_0\over8}
\left[{\rm Tr}\left(D_a\right)]^2+2{\rm Tr}\left(D_a^2\right)\right]
+
{\lambda_+\over8}
\left[{\rm Tr}\left(D_b\right)]^2+2{\rm Tr}\left(D_b^2\right)\right]
+{\lambda_H\over4}({\rm Tr}D_a)({\rm Tr}D_b)\;.
\eqa
Writing out explicitly the terms in $\Phi[D]$, we find
\bqa\nonumber
\Phi[D] &=& {3\over 8}\left[{\lambda_0}\sumint_KD_{11}\sumint_QD_{11}+
{\lambda_0}\sumint_KD_{22}\sumint_QD_{22}
+{\lambda_+}\sumint_KD_{33}\sumint_QD_{33}
+{\lambda_+}\sumint_KD_{44}\sumint_QD_{44}\right]
 \\ \nonumber & &  
+{1\over 4}\left[{\lambda_0}\sumint_KD_{11}\sumint_QD_{22}+{\lambda_+}
\sumint_KD_{33}\sumint_QD_{44}\right.
+{\lambda_H}\sumint_KD_{11}\sumint_QD_{33}
+{\lambda_H}\sumint_KD_{11}
\sumint_QD_{44}
\\ & &
+{\lambda_H}\sumint_KD_{22}\sumint_QD_{33}
\left.
+{\lambda_H}\sumint_KD_{22}\sumint_QD_{44} \right]\;.
\label{phi2exp}
\eqa
Again the terms involving the off-diagonal elements of the full propagator
$D$ are absent since they vanish upon summation over the Matsubara frequencies.

The gap equations now follow in the usual manner. They require 
renormalization, which is briefly discussed
in Appendix A. After renormalization, the gap equations
read
\bqa
\label{m12pi}
M^2_1 - m^2_1 &=& {1\over2}\left[
3\lambda_0({J}^c_1 + J^T_1) + \lambda_0({J}^c_2 + J^T_2) + 2\lambda_H({J}^c_3 + J^T_3)\right]
\;,\\
\label{m22pi}
M^2_2-m^2_2 &=& {1\over 2}\left[\lambda_0({J}^c_1 + J^T_1) + 3\lambda_0({J}^c_2 + J^T_2) + 2\lambda_H({J}^c_3 + J^T_3)\right]
\;,\\
M^2_3-m^2_3 &=& {1\over2}\left[
\lambda_H({J}^c_1 + J^T_1) + \lambda_H({J}^c_2 + J^T_2) + 4\lambda_+({J}^c_3 + J^T_3)\right]\;,\\
0 &=& \phi_0\left[m_0^2-\mu_0^2+{\lambda_0\over
2}\phi_0^2+{1\over2}\left[3\lambda_0(J^c_1+J^T_1) + \lambda_0(J^c_2+J^T_2) +
2\lambda_H(J^c_3+J^T_3)\right]\right]\;,
\label{phi2pi}
\eqa
In the appendix we argue that $M_4$ is equal to $M_3$
and their gap equations are idential.
This is only correct as long as there is no charged kaon condensate.
Substituting Eq.~(\ref{m12pi}) into Eq.~(\ref{phi2pi}), it can be written
as
\bqa
\phi_0\left[M^2_1-(\mu_0^2+{\lambda_0}\phi_0^2)\right] = 0\;,
\eqa
Similarly, substituting Eq.~(\ref{phi2pi}) into Eq.~(\ref{m22pi}), we obtain
\bqa
\label{gaphart}
M_2^2-\mu_0^2&=&
{\lambda_0}\left[(J_2^c+J_2^T)-(J_1^c+J_1^T)
\right]\;.
\eqa

The effective potential can be renormalized using the same methods and
after renormalization the effective potential is given by
\bqa\nonumber
\Omega&=&{1\over2}(m^2_0-\mu_0^2)\phi_0^2+{\lambda_0\over8}\phi_0^4
+ {1\over2}\left({\cal J}^c_1 + {\cal J}^T_1\right)
+ {1\over2}\left({\cal J}^c_2 + {\cal J}^T_2\right)
+ ({\cal J}^c_3 + {\cal J}^T_3)\\
\nonumber & & 
-\frac{1}{2}(M_1^2-m_1^2)({J}^c_1 + { J}^T_1)
-\frac{1}{2}(M_2^2-m_2^2)({ J}^c_2 + { J}^T_2)
-(M_3^2-m_3^2)({J}^c_3 + { J}^T_3)\\ \nonumber
&&
+ {3\lambda_0\over 8}
\left({J}^c_1 + J^T_1\right)^2
+ {3\lambda_0\over 8}\left({J}^c_2+J^T_2\right)^2
+\lambda_+\left({J}^c_3+J^T_3\right)^2 
+ {\lambda_0\over 4}
\left({J}_1^c +J^T_1\right)\left({J}_2^c +J^T_2\right)
\\ & & 
+{\lambda_H\over 2}
\left({J}^c_1+ J^T_1\right) \left({J}^c_2+ J^T_2\right) 
+ 
{\lambda_H\over 2}
\left({J}^c_2+ J^T_2\right)\left({J}^c_3+ J^T_3\right) \;.
\label{finalomega}
\eqa
\end{widetext}

\subsection{Phase diagram and quasiparticle masses}
In order to determine the phase diagram and numerical evaluate
the masses of the quasiparticle, we need to know the values of the
parameters in the effective Lagrangian~(\ref{lag2x2}).
In Ref.~\cite{alfordus}, the authors use a quark chemical potential 
$\mu\simeq 500$ MeV and and superconducting gap $\Delta\simeq 30$ MeV.
Extrapolating from asymptotically high densities, 
Eq.~(\ref{fasym}), and Eqs.~(\ref{pam1})--(\ref{pam4}) 
give $f_{\pi}\simeq100$ MeV, 
$\mu_i\simeq20$ MeV, $m_1\simeq5$ MeV, and $m_2\simeq4$ MeV.
In the numerical calculatations that we present below, we use 
$m_0=4$ MeV, $m_+=5$ MeV, $\mu_0=\mu_+=4.5$ MeV, and $f_{\pi}=100$ MeV  
unless otherwise stated. The renormalization scale is chosen to be the
average of the two vacuum masses, i.e. $\Lambda=4.5$ MeV.
Using Eqs.(\ref{pam5})--(\ref{paml}) and identifying 
$2\beta_1$ with $\lambda_0$, $2\beta_2$ with $\lambda_+$,
$2\alpha$ with $\lambda_H$, the numerical values are
$\lambda_0=1.25\times10^{-3}$, 
$\lambda_+=1,.08\times10^{-3}$, and $\lambda_H=1.16\times10^{-3}$.
In the plots, where we vary the chemical potentials, we use the same couplings
throughout as $\mu$-dependent coupling constants are problematic.

The neutral kaons $K^0$ and $\bar{K^0}$ 
are identified with the linear
combinations of the 
fields $\phi_1$ and $\phi_2$, and therefore with
$\tilde{\omega}_1(p)$ and $\tilde{\omega}_2(p)$. Similarly, the charged kaons
$K^-$ and $K^+$ are given by linear combinations of $\phi_3$ and $\phi_4$
and are identified with $\tilde{\omega}_3(p)$ and $\tilde{\omega}_4(p)$.

\begin{figure}[htb]
\begin{center}
  \includegraphics[width=7.3cm]{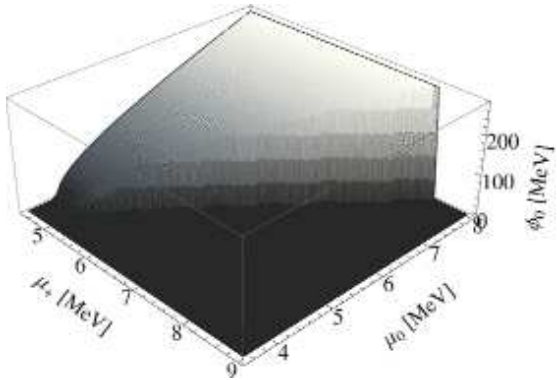}
\caption{Neutral
kaon condensate as a function of the chemical potentials
$\mu_0$ and $\mu_+$ for $T=0$.}
\label{cond1}
\end{center}
\end{figure}

In Fig.~\ref{cond1}, we show
the neutral kaon condensate as a function of $\mu_0$ and $\mu_+$ for $T=0$.
For $\mu_+=0$, i.e. along the $\mu_0$-axis, 
there is a second-order phase transition to a neutral phase with a
kaon condensate at a critical chemical potential $\mu_0=m_2$. 
This is the CFL-$K^0$ phase.
For larger values of $\mu_+$ the transition becomes first order.
The point in the $(\mu_0,\mu_+)$-plane where the transition changes order 
is a critical point and given by $(4.0,5.0)$ MeV.
In the part of the phase diagram where the transition is first order, the
transition is actually to a phase with 
condensate of charged kaons. This condensate is not
shown in the figure.
This is the CFL-$K^+$ phase.
Thus there is a competition between the neutral
and the charged condensates and nowhere do they exist simultaneously.
The transitions to the kaon-condensed phases are density driven.

\begin{figure}[htb]
\begin{center}
\includegraphics[width=7.3cm]{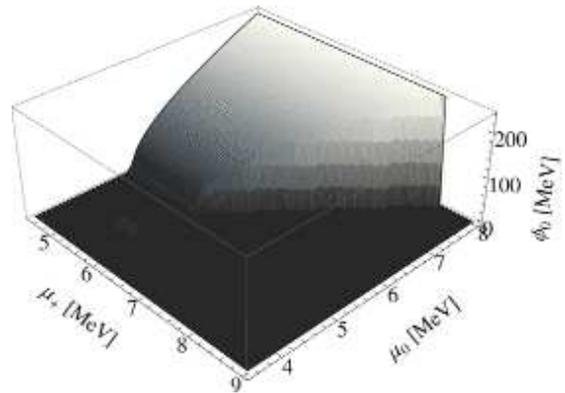}
\caption{Neutral 
kaon condensate as a function of the chemical potentials
$\mu_0$ and $\mu_+$ for $T=200$ MeV.}
\label{condensate}
\end{center}
\end{figure}

In Fig.~\ref{condensate}, we show the neutral
kaon condensate as a function of $\mu_0$ and $\mu_+$ for $T=200$ MeV.
The point in the $(\mu_0,\mu_+)$-plane where the transition changes order 
is given by $(5.2,6.0)$ MeV. 

\begin{figure}[htb]
\begin{center}
\includegraphics[width=7.3cm]{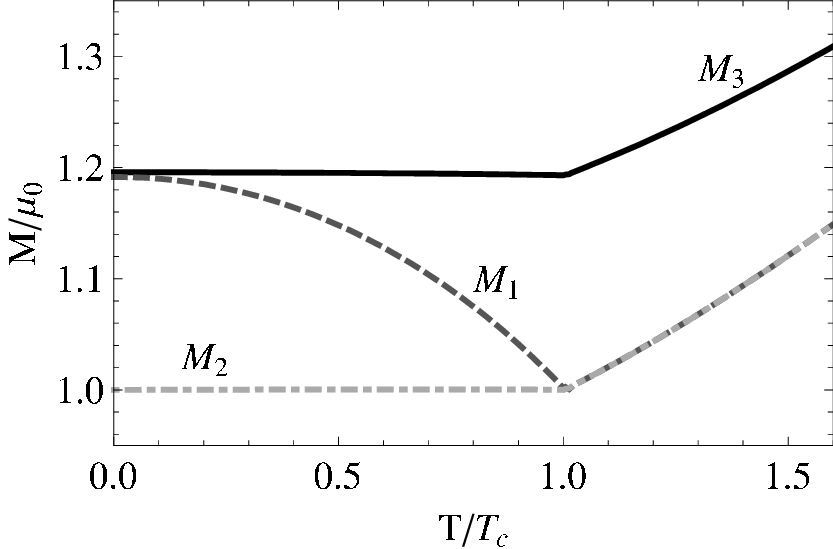}
\caption{Dressed masses $M_{1,2}$ and $M_3$ normalized to $\mu_0$
as a function of 
$T/T_c$ for $\mu_0=\mu_+=4.5$ MeV.}
\label{masses}
\end{center}
\end{figure}

In Fig.~\ref{masses}, we show the mass parameters $M_{1,2}$ and $M_{3}$
normalized to $\mu_0$
for $\mu_0=\mu_+=4.5$ MeV as a function of $T$ normalized to $T_c$. 
The masses $M_3$ and $M_4$ are degenerate for all values of $T$, while
$M_1$ and $M_2$ become degenerate at the critical temperarture.
If the Goldstone theorem is obeyed $M_2$ is exactly equal to $\mu_0$
in the broken phase. We notice that there is a tiny deviation.

\begin{figure}[htb]
\begin{center}
\includegraphics[width=7.3cm]{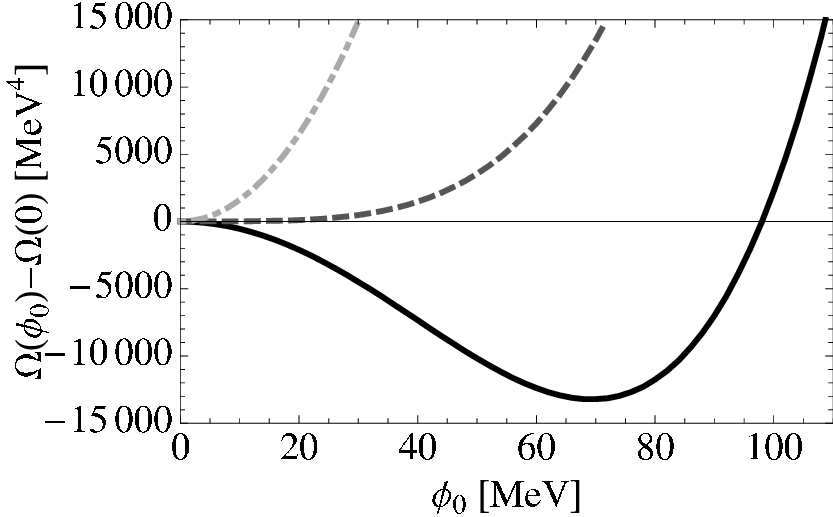}
\caption{$\Omega(\phi_0)-\Omega(0)$ for 
$\mu_0=\mu_+=4.5$ MeV and
three different values of the 
temperature.  The solid line is $T=0$, the dashed line is $T=T_c=118.5$ MeV,
and the dotted line is $T=200$ MeV.}
\label{vminv0}
\end{center}
\end{figure}

In Fig~\ref{vminv0}, we show the thermodynamic
potential $\Omega(\phi_0)-\Omega(0)$ as a function of the condensate
$\phi_0$ for $\mu_0=\mu_+=4.5$ MeV and
three different values of the temperature. 
$\Omega(\phi_0)$ is obtained by solving the gap equation for the masses
(\ref{m12pi}),~(\ref{m22pi}), and~(\ref{phi2pi}) and 
inserting the values into the effectifve potential~(\ref{finalomega}).
The solid line is $T=0$, the dashed line is $T=T_c=118.5$ MeV,
and the dotted line is $T=200$ MeV. The phase transition is second order.

\begin{figure}[htb]
\begin{center}
\includegraphics[width=7.3cm]{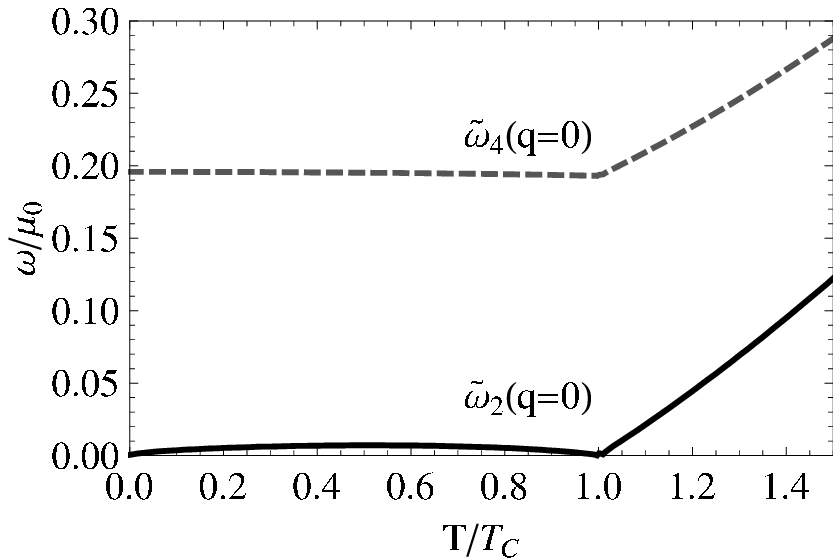}
\caption{Mass gaps for the $K^+$ and $K^0$ modes
for $\mu_0=\mu_+=4.5$ MeV and as a
function of $T$ normalized to $T_c$.}
\label{energygaps}
\end{center}
\end{figure}

In Fig.~\ref{energygaps} we show the masses of $K^0$ ($\omega_2(q=0)$) 
and $K^+$ ($\omega_4(q=0)$) for $\mu_0=\mu_+=4.5$  
MeV and as functions of $T$ normalized to $T_c$. We notice that the
mass of $K^0$ is not strictly zero, which  
explicitly shows that the Goldstone theorem is
not respected by the Hartree approximation~\footnote{Note that the mass gap
does vanish at $T=T_c$ since $J_1^c=J_2^c$ and $J_1^T=J_2^T$, 
cf. Eq~(\ref{gaphart}).}. In Ref.~\cite{alfordus}, the authors make
some further approximations of the sum-integrals appearing in the
gap equations. These approximations give rise to an exactly gapless mode.
As pointed out in their paper and as can be seen in 
Fig.~\ref{energygaps}, this is a very good approximation.

\begin{figure}[htb]
\begin{center}
\includegraphics[width=7.3cm]{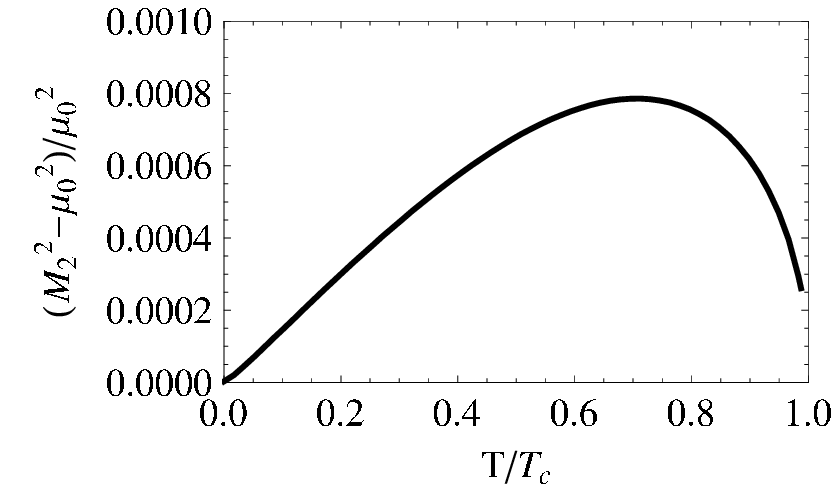}
\caption{Measure of the violation of Goldstone's theorem, 
$(M_2^2-\mu_0^2)/\mu_0^2$ as a function of $T/T_c$.}
\label{gbreak}
\end{center}
\end{figure}

In Fig.~\ref{gbreak}, we show the difference 
$M_2^2-\mu^2_0$, ~Eq.~(\ref{gaphart}), normalized to $\mu_0^2$ for $\mu_0=4.5$ 
MeV as 
a function of $T/T_c$. This is a another measure of the violation of Goldstone's
theorem. We see that the violation is tiny.

\subsection{Effects of charge neutrality}
Bulk matter in compact stars must be overall color and electrically neutral,
otherwise one pays an enormous energy penalty. The neutrality constraint
applies whether or not the gauge charges are broken or not.
For certain values of $\mu_0$, $\mu_+$, and $T$, the thermodynamically
stable state, i.e. $\Omega$ evaluated at the stationary points, has an overall
electric charge. For example, in the region of the 
lower corner of Fig.~\ref{cond1}, there is a charged kaon condensate which is 
positively charged.

In this subsection, we impose the constraint of overall charge neutrality.
In some case this constraint can have dramatic effects on the phase diagram.
For example, in a recent study by Abuki {\it et al}~\cite{abukku}, they
showed that pion condensation in the presence of finite
isospin chemical potential does not occur for physical masses of the
pions. In fact, they
found a tiny window of pion condensation for pion masses below 
approximately 10 KeV. Thus pion condensation is very sensitive to the
explicit chiral symmetry breaking.

\begin{figure}[htb]
\begin{center}
\includegraphics[width=7.3cm]{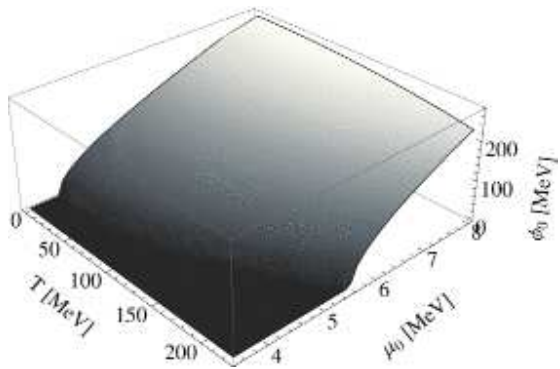}
\caption{The neutral kaon condensate as a function of $T$ and $\mu_0$
for a system which is electrically neutral.}
\label{neutralphase}
\end{center}
\end{figure}

For simplicity, we assume that we can describe the background of electrons
by an ideal Fermi gas. 
We then add to the Lagrangian~(\ref{lag2x2}), the term
\bqa
{\cal L}_{\rm electrons}&=&\tilde{\psi}_{e}\left(
\gamma^{\mu}\partial_{\mu}+\gamma^0\mu_ee-m_e
\right)\psi_{e}\;,
\eqa
where $\psi_e$ 
denotes the electron field, $e$ is the electron charge,
and $m_e$ is the mass of the electron.
The contribution to the
free energy from the electrons is denoted by $\Omega_e$ and reads
\bqa\nonumber
\Omega_e&=&-2\int_p\left\{E_p
+T\log\left[1-e^{-\beta(E_p-\mu_Q)}\right]
\right.\\ &&
\left.
+T\log\left[1-e^{-\beta(E_p+\mu_Q)}\right]\right\}\;.
\label{massel}
\eqa
where $E_p=\sqrt{p^2+m^2_e}$. In the following, we assume the electrons are 
massless. If we use dimensional regularization, the first term in 
Eq.~(\ref{massel}) vanishes since there is no mass scale.
The temperature-dependent integrals can be done analytically and we obtain
\bqa
\label{eladd}
\Omega_e&=&-{1\over12\pi^2}\mu_Q^4-{1\over6}\mu_Q^2T^2
-{7\pi^2\over180}T^4\;.
\eqa
We add Eq.~(\ref{eladd}) to Eq.~(\ref{lag2x2}) to obtain the full thermodynamic
potential. The contribution $n_+$ to the electric charge density 
from the kaons is then given by
\bqa\nonumber
n_+&=&
-{\partial \Omega\over\partial\mu_Q} \\
&=&-{1\over2}{\rm Tr}\left[
{\partial D_0^{-1}\over \partial\mu_+}D
\right]\;.
\label{n+}
\eqa
Using 
Eqs.~(\ref{finalomega}),~(\ref{eladd}), and~(\ref{n+}), we obtain
\bqa
n_+
&=&
2\sumint_Q{\mu_+(-\omega_n^2+p^2+M_3^2-\mu_+^2)
\over[\omega_n^2+\tilde{\omega}_3^2(q)][\omega_n^2+\tilde{\omega}_4^2(q)]}\;.
\eqa
This expression is free of ultraviolet divergences. 
After summing over Matsubara frequencies
and averaging over angles, the equation reduces to 
\bqa\nonumber
n_+&=&{1\over4\pi^2}\int_0^{\infty}\left[
{1\over e^{\tilde{\omega}_3(q)/T}-1}-{1\over e^{\tilde{\omega}_4(q)/T}-1}
\right]q^2\,dq\;.
\\ &&
\eqa
The contribution $n_e$ to the electric charge density from the electrons
is given by
\bqa\nonumber
n_e&=&
-{\partial \Omega_e\over\partial\mu_Q} \\
&=&
-{1\over3\pi^2}\mu_Q^3-{1\over3}\mu_QT^2\;.
\eqa
Charge neutrality amounts to requiring that 
\bqa
n_++n_e&=&0\;.
\label{charge0}
\eqa
In the case where we do not impose the charge neutrality 
requirement~(\ref{charge0}), we have two chemical potentials that we can vary
freely. The charge neutrality constraint gives a relation between $\mu_0$
and $\mu_+$ and only one of them is free to vary.
For a given value of $T$ and e.g. $\mu_0$, we must therefore
solve simultaneously the gap equations~(\ref{m12pi})--(\ref{gaphart})
and the equation~(\ref{charge0}), to find the dressed masses $M_i$, $\phi_0$
and the chemical potential $\mu_+$.
In Fig.~\ref{neutralphase}, we show the neutral kaon condensate
for neutral matter as a function of $T$ and $\mu_0$
For $T=0$, the onset of kaon condensation is at $\mu_0=m_0$ as expected.
The transition is second order for all values of $\mu_0$ and again 
the transition is to a symmetric phase.

\begin{figure}[htb]
\begin{center}
\includegraphics[width=7.3cm]{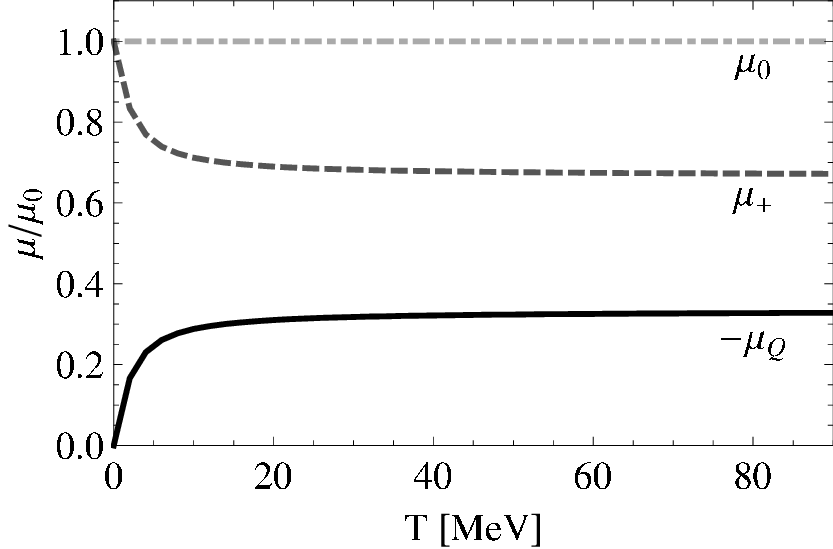}
\caption{The chemical potentials
$\mu_+$ and $-\mu_Q$ divided by $\mu_0$ as functions of temperature
in an electrically neutral system.}
\label{muqmu+}
\end{center}
\end{figure}

In Fig.~\ref{muqmu+}, we show $\mu_+$ and $-\mu_Q=\mu_0-\mu_+$ 
normalized to $\mu_0$ as functions
of temperature $T$. The sum of the curves are always equal to one.
The chemical potential $-\mu_Q$ vanishes at $T=0$, increases rapidly, and is
essentially constant for $T\geq20$ MeV.
The chemical potential $\mu_+$, which equals $\mu_0$ at $T=0$,
never reaches its critical value 
$\mu_+^c=m_+$ and this explains why the phase transition in 
Fig.~\ref{neutralphase} is to a symmetric phase and not to a charged kaon
condensed phase. In other words, charge neutrality does not allow for
a charged condensate.

Finally we compare the critical temperature with and without neutrality.
In the case with no neutrality constraint, the critical temperature was 
$118.5$ MeV for $\mu_0=\mu_+=4.5$ MeV. In the neutral case, $\mu_+$
is a function of $\mu_0$ and so varies with temperature. At $T=0$, we
always have $\mu_Q=0$, i. e. $\mu_0=\mu_+$. We therefore choose the same value,
i.e.
$\mu_0=4.5$ MeV. The critical temperaature in this case is $T_c=125$ MeV.
In other words, there is a small increase in the critical temperature
as one impose charge neutrality. This is in agreement with the 
findings of Ref.~\cite{alfordus}.

\section{Summary and Outlook}
In the present paper, we have studied the $O(N)$
and $O(2)\times O(2)$ models with chemical potentials using the 2PI
formalism in the 
Hartree approximation. We have explicitly
shown it is possible to renormalize the
gap equation and effective potential in a way that is independent 
of temperature and chemical potentials. We have studied the phase
diagram and the quasiparticle masses of the $O(2)\times O(2)$ model 
where the neutral kaons condense at sufficiently low temperature and
sufficiently large value of the chemical potential.
If the transition is first order, it turns out that the transition is not
to a symmetric state but to a state with a $K^+$ condensate.
This is in agreement with the findings of Alford, Braby,
and Schmitt~\cite{alfordus}.
Generally, our predictions for the critical temperature and other quantities
are different from theirs
since our couplings do not depend on the
chemical potentials. Secondly, we have not made a high-temperature approximation
of the thermal integrals in the gap equations that determine the dressed masses.
Finally, our gap equations and 
effective potential include renormalization effects.

One drawback of the 2PI Hartree approximation is that it does not
obey Goldstone's theorem. We have shown this explicitly 
and quantified the deviation by the right-hand-side of Eq.~(\ref{gaphart}).
For practical purposes, the 
violation is negligible which is reassuring. 
We therefore believe that the 2PI Hartree approximation is a useful
nonperturbative approximation for systems in thermal equilibrium.



The Hartree approximation and the large-$N$ limit are both mean-field
approximations. It would be desireable to go beyond mean field
for example by including next-to-leading corrections in the $1/N$-expansion.
In the context of pions, this has been done using the 1PI $1/N$-expansion,
but only for vanishing chemical potential~\cite{abw,jenstomas}.
Another method is the functional renormalization 
group method, see Refs.~\cite{wett1,wett2,gies,delamotte} and 
references therein.
The functional renormalization group method is a nonperturbative approach
that has been very succesful in finite-temperature field theory.
The essence of this approach is a flow equation for the average
effective action. This flow equation cannot be solved exactly, but one
must resort to approximations. A systematic approximation is the derivative
expansion of the effective action and even the simplest truncation, namely
the local-potential approximation, often yields good results.
In the context of Bose condensation, functional 
renormalization group methods have been applied to the nonrelativistic
case in Refs.~\cite{jensmike,jens2,cw,jp}. 
Work on pion and kaon condensation using  these methods
is in progress~\cite{jensleg}.

\appendix
\section{Renormalization}
In this section, we first discuss the renormalization of the gap equations
and effective potential for the $O(2N)$-symmetric models in detail.
We then briefly skecth the renormalization of the 
gap equations of the $O(2)\times O(2)$-symmetric model.
\subsection{$O(2N)$-symmetric models}

The effective potential and the gap equations contain divergent sum-integrals. 
For the theory to be renormalizable, the 
divergent terms must be independent of the 
the temperature $T$ and the chemical potential $\mu$.
Below we show explicitly that there are individual $\mu$-dependent
contributions 
to the gap equations which are divergent, but they cancel amongst themselves
and so the counterterms are the same as those needed to renormalize the
theory in the vacuum. 

There is a complication regarding the renormalization of the quartic 
coupling in the 2PI Hartree approximation. The truncatation of the
2PI effective action allows one to define two independent four-point 
functions and thus two independent sets of counterterms associated with 
them~\cite{berges2}. They correspond to the two $O(N)$ invariants 
$\delta_{ij}\delta_{kl}$ and 
$\delta_{ik}\delta_{jl}+\delta_{ik}\delta_{jk}$ from which 
$F_{ijkl}$ is built. More generally, each $O(N)$-invariant term in the
2PI effective action has its own independent counterterm.
For example, the counterterms  arising from the classical potential
in Eq.~(\ref{2piea}) are written as 
\bqa
{1\over2}\delta m^2_0+{\delta\lambda_0\over8N}\phi_0^4\;,
\eqa
while the counterterms arising from the term ${\rm Tr}D_0^{-1}D$ are
written as 
\bqa
\label{countd01}
\delta m_1^2&=&\delta m^2
+{\delta \lambda_2^A+2\delta\lambda_2^B\over2N}\phi_0\;,
\\
\delta m_2^2&=&\delta m^2
+{\delta \lambda_2^A\over2N}\phi_0\;.
\label{countd02}
\eqa
The two counterterms 
and $\delta\lambda^A$ and $\delta\lambda^B$.
correspond to the two $O(N)$-invariant terms
${\rm Tr}[\phi^2_0]{\rm Tr}[D]$ and ${\rm Tr}[\phi^2_0D]$ 
in the expression for ${\rm Tr}D_0^{-1}D$ 
that scale as $N$ and one, respectively.
It was shown in Ref.~\cite{fejos} that $\delta m_0^2=\delta m^2$
and $\delta\lambda_0=\delta\lambda^A+2\lambda^B$.

Renormalization of the gap equations can be done by applying
an iterative procedure which is discussed in Refs.~\cite{bir3,fejos}.
To this end, we write the mass counterterm 
and the coupling constant counterterms
$\delta m^2$, $\delta\lambda^A$, and $\delta\lambda^B$
as power series in $\lambda$:
\bqa
\delta m^2 &=&
\sum_{n=1}^{\infty}\delta m_n^2 \label{eq:counterm} \;,\\
\delta\lambda^A&=&
\sum_{n=1}^{\infty}\delta \lambda_n^A \;,\\
\delta\lambda^B&=&
\sum_{n=1}^{\infty}\delta\lambda_n^B\;, \label{eq:counterB}
\eqa
where $\delta m_n^²$ is a counterterm of order $\lambda^n$, and
$\delta\lambda_n^A$ and $\delta\lambda_n^B$ are counterterms of order
$\lambda^{n+1}$, respectively. Similarly, we write the self-energies 
$\Pi_{i}$ as power series in $\lambda$:
\bqa
\Pi_{i}&=& \Pi_{i}^{(1)} + \Pi_{i}^{(2)} + \cdots\;,
\label{eq:Pipower}
\eqa
where the superscript indicates the power of $\lambda$. Inserting 
Eqs.~(\ref{eq:counterm})--(\ref{eq:Pipower})
into (\ref{eq:D}), we can determine $\Pi_n^{(n)}$, $\delta m^2_n$,
$\delta\lambda_n^A$, and $\delta\lambda_n^B$ by iteration.
\begin{widetext}
The Schwinger-Dyson equation (\ref{eq:D}) for the three diagonal components
of the propagator is
\bqa
\label{scwdys1}
D^{-1}_{11}&=& \left(D_0^{-1}\right)_{11}
+{\lambda\over2N}\left[
3\sumint_Q D_{11} + \sumint_Q D_{22} + (2N-2)\sumint_Q D_{33}\right]
\;,\\
\label{scwdys2}
D^{-1}_{22}&=& \left(D_0^{-1}\right)_{22}
+{\lambda\over2N}\left[
\sumint_Q D_{11} + 3\sumint_Q D_{22} + (2N-2)\sumint_Q D_{33}\right]
\;,\\
D^{-1}_{33}&=& \left(D_0^{-1}\right)_{33}
+{\lambda\over2N}\left[
\sumint_Q D_{11} + \sumint_Q D_{22} + 2N\sumint_Q D_{33}\right]
\label{scwdys3}
\;.
\eqa
Since the off-diagonal parts of the self-energy 
vanish we have 
$(D_0)_{ij}^{-1}=(D)_{ij}^{-1}$ for $i\neq j$.
The first iteration of the gap equation is found by ignoring the self-energy
$\Pi$
on the right-hand side of the gap equations~(\ref{scwdys1})--(\ref{scwdys3}),
i.e. one replaces the full propagator by the tree-level propagator.
If we denote the components of the free propagator by $(D_0)_{ij}$,
we can write
\bqa
\label{gapm1}
M^2_1 - m^2_1 &=& \delta m^2+{\delta\lambda_1^A + 2\delta\lambda_1^B\over2N}
\phi_0^2
+{\lambda\over2N}\left[
3\sumint_Q (D_0)_{11} + \sumint_Q (D_0)_{22} + (2N-2)\sumint_Q (D_0)_{33}\right]
\;,\\
M^2_2-m^2_2 &=& \delta m^2+{\delta\lambda_1^A\over2N}\phi_0^2
+{\lambda\over2N}\left[
\sumint_Q (D_0)_{11} + 3\sumint_Q (D_0)_{22} + (2N-2)\sumint_Q (D_0)_{33}\right]
\;,\\
M^2_3-m^2_3 &=& \delta m^2+{\delta\lambda_1^A\over2N}\phi_0^2
+{\lambda\over2N}\left[
\sumint_Q (D_0)_{11} + \sumint_Q (D_0)_{22} + 2N\sumint_Q (D_0)_{33}\right]
\;,
\label{gapp33}
\eqa
\end{widetext}
where we have used Eqs.~(\ref{countd01})--(\ref{countd02}).
The diagrammatic interpretation of the procedure is shown in 
Fig.~\ref{hartree22}. The propagator in the loops are free propagators
and the first Feynman diagram on the right-hand side corresponds to the
sum-integrals in Eqs.~(\ref{gapm1})--(\ref{gapp33}).
We now consider in detail the renormalization of Eq.~(\ref{gapm1}).
The sum-integrals can be split into divergent temperature-independent
and temperature-dependent convergent parts and so we write
\bqa
\sumint_Q (D_0)_{nn} = I^d_{n} + I^T_{n} \;. 
\label{eq:split}
\eqa
The free propagator $D_0$ is obtained from Eq.~(\ref{fulld}) by replacing the
medium-dependent masses $M_i$ by $m_i$. The first sum-integral
in Eq.~(\ref{gapm1}) 
then becomes
\bqa
\sumint_Q (D_0)_{11} = 
{{\omega_n^2+p^2+m_1^2-\mu^2}
\over(\omega_n^2+\omega_1^2)(\omega_n^2+\omega_2^2)}
\;.
\label{d11sum}
\eqa

\begin{widetext}
We begin by summing over the Matsubara frequences. This yields
\bqa
\sumint_Q (D_0)_{11} &=& \frac{1}{4} \int_q  \left[ {1\over\omega_1}\left(1+
{{1\over 2}(m_1^2-m_2^2)+2\mu^2\over \sqrt{4\mu^2
\left[q^2+\frac{1}{2}(m_1^2+m_2^2)\right] + 
\frac{1}{4}(m_1^2-m_2^2)^2}}\right)\coth
\left({\omega_1\over 2T}\right) \right. \nonumber\\
&& \left. + {1\over\omega_2}\left(1-
{{1\over 2}(m_1^2-m_2^2)+2\mu^2\over \sqrt{4\mu^2
\left[q^2+\frac{1}{2}(m_1^2+m_2^2)\right] + 
\frac{1}{4}(m_1^2-m_2^2)^2}}\right)\coth\left({\omega_2\over 2T}\right)
\right]\;.
\label{sumint11}
\eqa

We are interested in isolating the divergent parts of the sum-integral.
Dropping the convergent temperature-dependent parts of
$\coth\left({\omega_{1,2}\over 2T}\right)$ and rearranging, we obtain
\bqa\label{ymse}  
I^d_{1} &= &
\frac{1}{4} \int_q \left[\left({1\over\omega_1}+{1\over\omega_2}\right)
+\left({1\over\omega_1}-{1\over\omega_2}\right)
{{1\over 2}(m_1^2-m_2^2)+2\mu^2\over \sqrt{4\mu^2
\left[q^2+\frac{1}{2}(m_1^2+m_2^2)\right]+\frac{1}{4}(m_1^2-m_2^2)^2}}\right] 
\;.
\eqa
For convenience, we introduce some shorthand notation:
\bqa
E_1 &\equiv & \sqrt{q^2 + m_1^2} \;,\\
A &\equiv &-{1\over 2}\left(m^2_1-m^2_2\right)+\mu^2 \;,\\
B &\equiv &\sqrt{4\mu^2\left[q^2+\frac{1}{2}(m_1^2+m_2^2)\right] 
+ \frac{1}{4}(m_1^2-m_2^2)^2} \;.
\eqa
The various factors in Eq.~(\ref{ymse}) can then be compactly
written as
\bqa
\omega_{1,2}(q) &=& E_1\sqrt{1+{A\pm B\over E_1^2}}\;,\\
{{1\over 2}(m_1^2-m_2^2)+2\mu^2\over \sqrt{4\mu^2
\left[q^2+\frac{1}{2}(m_1^2+m_2^2)\right] + \frac{1}{4}(m_1^2-m_2^2)^2}} 
&=& {-A+3\mu^2\over B}\;.
\eqa
The next step is to expand $\omega_{1,2}(q)$ in inverse
powers of $E$. The integral $I_d^1$ can then be written as
\bqa
I^d_{1}&=& {1\over 4}\int_q \left\{{1\over E_1}
\left[\left(1 - {A-B\over 2E_1^2} + {3(A-B)^2\over 8E_1^4}+\cdots\right) + 
\left(1 - {A+B\over 2E_1^2} + {3(A+B)^2\over 8E_1^4}+\cdots\right)\right] 
\right.\nonumber \\ \nonumber
&&\left.- {1\over E_1}\left[\left(1 - {A-B\over 2E_1^2} + 
{3(A-B)^2\over 8E_1^4}+
\cdots\right) - \left(1 - {A+B\over 2E_1^2} + 
{3(A+B)^2\over 8E_1^4}+\cdots\right)\right]{-A+3\mu^2\over B}\right\} \\
&=&
{1\over 4}\int_q 
\left[{2\over E_1}-{\mu^2(m_1^2-m_2^2)\over2E_1^5} 
+ \cdots \right]\;.
\eqa
\end{widetext}
We note that the integrals of the form $\int_q E_1^{-n}$ are 
divergent in the ultraviolet
for $n \leq 3$ and convergent for $n>3$. Since the terms with $n=3$ cancel
in the integral, there is only one divergent term. We denote this term
by $I$ and it reads
\bqa
I&=&{1\over2}\int_q{1\over\sqrt{q^2+m^2_1}}\;.
\eqa
This integral is easily calculated with dimensional regularization
\bqa
I
&=&-{m^2_1\over(4\pi)^2}\left({\Lambda^2\over m^2_1}\right)^\epsilon
\left[{1\over\epsilon}+1+{\cal O}(\epsilon)\right]\;.
\eqa
Having isolated the UV divergence, the temperature-independenet part of the
sum-integral in Eq.~(\ref{sumint11}) can now be written as
\bqa
I_1^d&=&-{m_1^2\over(4\pi)^2\epsilon}+
I_1^c+{\cal O}(\epsilon)
\;,
\eqa
where $I_1^c$ is the finite part of $I_1^d$ and is given by
\begin{widetext}
\bqa\nonumber
I^c_{1} &=& 
\frac{1}{4} \int_q \left\{\left[{1\over\omega_1}
+{1\over\omega_2}\right]
+\left[{1\over\omega_1}-{1\over\omega_2}\right]{{1\over 2}(m_1^2-m_2^2)
+2\mu^2\over \sqrt{4\mu^2\left[q^2+\frac{1}{2}(m_1^2+m_2^2)\right] + 
\frac{1}{4}(m_1^2-m_2^2)^2}}  - {2\over \sqrt{q^2 + m_{1}^2}}\right\}
-{m_1^2\over(4\pi)^2}\left(L_1+1\right)
\;,
\\
\label{ic12}
\eqa
where $L_n = \log{\Lambda^2\over m_n^2}$.
The temperature-dependent part of the sum-integral~(\ref{sumint11}) is given by
\bqa
\label{it12}
I^T_{1} &=& \frac{1}{2} \int_q 
\left\{\left[{1\over\omega_1(e^{\omega_1/T}-1)}
+{1\over\omega_2(e^{\omega_2/T}-1)}\right]\right.\nonumber\\
& & \left.-\left[{1\over\omega_1(e^{\omega_1/T}-1)}
+{1\over\omega_2(e^{\omega_2/T}-1)}\right]{{1\over 2}
(m_1^2-m_2^2)+2\mu^2\over \sqrt{4\mu^2
\left[q^2+\frac{1}{2}(m_1^2+m_2^2)\right] + 
\frac{1}{4}(m_1^2-m_2^2)^2}}\right\}\;.
\eqa
The integrals in Eqs.~(\ref{ic12})--(\ref{it12}) 
are now finite in three dimensions and
we set $\epsilon=0$. The contributions to the right-hand side of 
Eq.~(\ref{gapm1}) from $(D_0)_{22}$ and $(D_0)_{33}$
are calculated in the same manner.
The gap equation can then be written as
\bqa\nonumber
M_1^2-m^2_1&=&\delta m^2
+{\delta\lambda_1^A\phi_0^2+2\delta\lambda_1^B\over2N}
\phi_0^2
+{\lambda\over2N}\bigg\{
\left[3m_1^2+m_2^2+(2N-2)m_3^2\right]{1\over\epsilon}
+3I_1^c+I_2^c+(2N-2)I_3^c
.\\ &&
+3I_1^T+I_2^T+(2N-2)I_3^T
\bigg\}\;.
\label{divgap}
\eqa
\end{widetext}
where $I_2^c$ and $I_2^T$ are obtained by exchanging $m_1^2$ for $m_2^2$ and vice versa
in Eqs.~(\ref{ic12}) and~(\ref{it12}), and
\bqa
I^c_{3} &=& -{m_3^2\over(4\pi)^2}\left(L_3+1\right),
\eqa
\bqa
I^T_3&=& \frac{1}{2\pi^2} \int^\infty_0 
{\mathrm{d}q q^2\over \sqrt{q^2+m_3^2}}{1\over e^{\sqrt{q^2+m_3^2}/T}-1}\;.
\label{it3}
\eqa
The divergences in Eq.~(\ref{divgap}) are now cancelled by choosing the
counterterms appropriately. This gives 
\bqa
\label{count1}
\delta m_1^2&=&{\lambda \over(4\pi)^2\epsilon}\left(1+{1\over N}\right)m^2\;,\\
\label{count2}
\delta\lambda_1^A&=&{\lambda^2\over(4\pi)^2\epsilon}
\left(1+{2\over N}\right)\;,\\
\label{count3}
\delta\lambda_1^B&=&{\lambda^2\over(4\pi)^2\epsilon}{1\over N}\;.
\eqa
After renormalization, the first iteration of the 
gap equation for $M_1$ can be written as
\bqa\nonumber
M_1^2&=&m^2_1
+{\lambda\over2N}\left[
3I_1^c+I_2^c+(2N-2)I_3^c
\right.\\ &&
\left.
3I_1^T+I_2^T+(2N-2)I_3^T
\right]\;.
\eqa
The procedure is carried out iteratively to all orders in 
$m_n^2$ and $\lambda$. 
The final result is given in Eqs.~(\ref{m1dress1})-(\ref{m3dress3})
The counterterms $\delta m_n^2$
$\delta \lambda_n^A$, and $\delta \lambda_n^B$ are then expressed in terms
of $\delta m_{n-1}^2$, $\delta\lambda_{n-1}^A$, and 
$\lambda_{n-1}^B$ and obtain the recursion relations 
\bqa
\delta m_n^2&=&{\lambda \over(4\pi)^2\epsilon}\left(1+{1\over N}\right)
m^2_{n-1}\;,\\
\delta\lambda^A_n&=&{\lambda \over(4\pi)^2\epsilon}
\left[\left(1+{1\over N}\right)
\delta\lambda^A_{n-1}+{\delta\lambda^B_{n-1}\over N}\right]
\;,\\
\delta\lambda^B_n&=&{\lambda \over(4\pi)^2\epsilon}
{\delta\lambda^B_{n-1}\over N}\;.
\eqa
The counterterms are in agreement with those found by Fejos, Patkos, and
Szep~\cite{fejos}. 
We note in particular that they are independent of 
temperature and 
chemical potential. 
They simplify significantly in the large-$N$ limit, where 
$\delta\lambda^B=0$.
Renormalizing to all orders effectively means that 
we replace the tree-level masses $m_i$ on the right-hand side of the gap 
equations by the medium dependent masses $M_i$.
For each $I_n$ in Eqs.~(\ref{ic12}),~(\ref{it12}), and~(\ref{it3})  we 
therefore define a 
corresponding $J_n$, where the tree level masses $m_n$ are replaced by the 
dressed masses $M_n$. The final result is given in 
Eqs.~(\ref{m1dress1})--(\ref{m3dress3}).

\begin{widetext}
We next consider the gap equation~(\ref{former}). Differentiating $\Omega$
with respect to $\phi_0$, we obtain unrenormalized gap equation
\bqa
0&=&\phi_0\left[
m^2-\mu^2+{\lambda\over2N}\phi_0^2
+{\lambda\over2N}\left(3\sumint_QD_{11}
+\sumint_QD_{22}
+(2N-2)\sumint_QD_{33}
\right)
\right]\;.
\label{urgap}
\eqa
Renormalizing the gap equation~(\ref{urgap}) in the same way as above,
we obtain
\bqa
0 &=& \phi_0\left[m^2-\mu^2+{\lambda\over
2N}\phi_0^2+{\lambda\over2N}\left[3(J^c_1+J^T_1) + (J^c_2+J^T_2) +
(2N-2)(J^c_3+J^T_3)\right]\right]\;.
\eqa

We should make sure that the counterterms that renormalize
the gap equations are also sufficient to 
renormalize the effective potential~(\ref{2piea}). 
We next consider separately the different terms contributing to $\Omega$. 
The first term is
${\rm Tr} \log  D^{-1}$. 
Taking the trace yields
\bqa
\frac{1}{2}{\rm Tr}\log D^{-1} &=&
{1\over2}
\sumint_Q\log\left[\omega_n^2+\tilde{\omega}_1^2(q)\right] 
+{1\over2}
\sumint_Q\log\left[\omega_n^2+\tilde{\omega}_2^2(q)\right]  
+(N-1)\sumint_Q\log\left[\omega_n^2+\tilde{\omega}_3^2(q)\right]\;.
\label{eq:trlog}
\eqa
To carry out renormalization by the iterative procedure, we expand
the sum-integrals in powers of the self-energies $\Pi_i$.
To first order in the self-energies, we obtain
\bqa\nonumber
\frac{1}{2}{\rm Tr}\log D^{-1} &=&
{1\over2}
\sumint_Q\log\left[\omega_n^2+{\omega}_1^2(q)\right] 
+{1\over2}
\sumint_Q\log\left[\omega_n^2+{\omega}_2^2(q)\right]  
+(N-1)\sumint_Q\log\left[\omega_n^2+{\omega}_3^2(q)\right]
\nonumber\\
&&
\hspace{-1.0cm}
+
{1\over2}\Pi_1^{(1)}
\sumint_Q{\omega_n^2+q^2+m_1^2+\mu^2\over
(\omega_n^2+\omega_1^2)(\omega_n^2+\omega_2^2)}
+{1\over2}\Pi_2^{(1)}
\sumint_Q{\omega_n^2+q^2+m_2^2+\mu^2\over
(\omega_n^2+\omega_2^2)(\omega_n^2+\omega_2^2)}
+(N-1)\Pi_3^{(1)}\sumint_Q{1\over\omega_n^2+\omega_3^2}\;.
\label{expansie}
\eqa
Summing over the Matsubara frequencies, each term involving 
$\log(\omega_n^2+\omega_{i}^2(q))$
can be written in the form
\bqa
\sumint_Q\log\left[\omega_n^2+\omega_{i}^2(q)\right] &=& 
{\cal I}_n^d+{\cal I}_n^T\;,
\eqa
where
\bqa
{\cal I}_n^d&=&  \int_q\,\omega_{i} (q)
\label{trlog1}
\\ 
{\cal I}_n^c&=&2T\int_q\log\left[1-e^{-\omega_{i}(q)/T}\right]\;.
\eqa
We see right away that the temperature-dependent integrals are  
convergent. 
We now consider the temperature-independent 
term in Eq.~(\ref{trlog1}) where $n=1,2$.
With the expected result in mind we rewrite $\omega_1+ 
\omega_2 = (\omega_1+\omega_2)/2+(\omega_1+\omega_2)/2$ and expand the  
terms in the first parantheses in $E_1 = \sqrt{q^2+m_1^2}$ and the  
second in $E_2 = \sqrt{q^2+m_2^2}$. This yields
\bqa
\omega_1(q)+\omega_2(q)= 
E_1+E_2 - \frac{1}{4}(m_1^2-m_2^2)\left(\frac{1}{E_1}-\frac{1} 
{E_2}\right)
-\frac{1}{16}(m_1^2-m_2^2)^2\left(\frac{1}{E_1^3}+\frac{1} 
{E_2^3}\right)
+\mathcal{O}(1/E_1^5)+\mathcal{O}(1/E_2^5)
\label{divtrlog}
\;.
\eqa
The first two terms are the divergences we get when $\mu = 0$. As for  
the other terms, writing $E_2^2 = E_1^2-(m_1^2-m_2^2)$
and expanding the resulting expression in $E_1$, we find that the  
divergent parts do in fact cancel:
\bqa
  - \frac{1}{4}(m_1^2-m_2^2)\left(\frac{1}{E_1}-\frac{1}{E_2}\right)
-\frac{1}{16}(m_1^2-m_2^2)^2\left(\frac{1}{E_1^3}+\frac{1} 
{E_2^3}\right)
+\mathcal{O}(1/E_1^5)+\mathcal{O}(1/E_2^5)
= \mathcal{O'}(1/E_1^5)\;.
\eqa
\end{widetext}
This shows that the divergent term in each sum-integral is on the form
\bqa
K_n&=&\int_q\sqrt{q^2+m_n^2}\;.
\eqa
With dimensional regularization, the integral can be easily calculated
and reads
\bqa
K_n&=&-{m_n^4\over2(4\pi)^2}
\left({\Lambda^2\over m_n^2}\right)^{\epsilon}
\left[{1\over\epsilon}+{3\over2}\right]\;.
\eqa
\begin{widetext}
To zeroth order in the self-energies $\Pi_i$, we can write
\bqa
{1\over2}{\rm Tr}\log D^{-1}&=&
-{1\over4(4\pi)^2}\left[m_1^4+m_2^4+2(N-1)m_3^2\right]{1\over\epsilon}
+{1\over 2}({\cal I}_1^c+{\cal I}_1^T)
+{1\over 2}({\cal I}_2^c+{\cal I}_2^T)
+(N-1)({\cal I}_3^c+{\cal I}_3^T)\;,
\label{zeroth}
\eqa
where
\bqa
{\cal I}_{1,2}^c&=&\int_q\left[\omega_{1,2}(q)-E_{1,2}\right]
-{m_n^4\over2(4\pi)^2}\left[L_n+{3\over2}\right]
\;,
\\ 
{\cal I}_3^c&=&
-{m_n^4\over2(4\pi)^2}
\left[L_3+{3\over2}\right]\;.
\eqa
We next consider the term ${\rm Tr}D_0^{-1}D$. Using that $\Pi=D^{-1}-D_0^{-1}$,
we obtain ${\rm Tr}D_0^{-1}D={\rm Tr}I-{\rm Tr}\Pi D$, where
$I$ is the identity matrix. The first term vanishes in dimensional
regularization and we are left with ${\rm Tr}\Pi D$.
Since $\Pi$ is diagonal, we obtain
\bqa
{1\over2}{\rm Tr}\Pi D
&=&{1\over2}\Pi_1
\sumint_Q{\omega_n^2+q^2+M_1^2+\mu^2\over
(\omega_n^2+\tilde{\omega}_1^2)(\omega_n^2+\tilde{\omega}_2^2)}
+{1\over2}\Pi_2
\sumint_Q{\omega_n^2+q^2+M_2^2+\mu^2\over
(\omega_n^2+\tilde{\omega}_2^2)(\omega_n^2+\tilde{\omega}_2^2)}
+(N-1)\Pi_3\sumint_Q{1\over\omega_n^2+\tilde{\omega}_3^2}\;.
\label{piddd}
\eqa
Since we are expanding the self-energies as $\Pi_i=\Pi_i^{(1)}+\Pi_i^{(2)}+...$,
the leading term is obtained by replacing $\Pi_i$ by $\Pi_i^{(1)}$ and
$\tilde{\omega_i}$ by ${\omega_i}$. We then see that the leading term
exactly cancels against the terms on the second line in Eq.~(\ref{expansie}).
At higher orders this no longer the case. 
The term ${\rm Tr}\Pi D$ then gives rise to an additional 
term which after renormalization reads
\bqa
-\frac{1}{2}(M_1^2-m_1^2)({J}^c_1 + { J}^T_1)
-\frac{1}{2}(M_2^2-m_2^2)({ J}^c_2 + { J}^T_2)
-(N-1)(M_3^2-m_3^2)({J}^c_3 + { J}^T_3)\;.
\eqa
Finally, we consider the two-loop diagrams in $\Phi[D]$.
The terms are all given by products of one-loop sum-integrals that we
have already calculated
\bqa
\Phi[D]&=&{\lambda\over8N}
\left\{
\left[\sumint_QD_{11}+\sumint_QD_{22}+2(N-1)\sumint_QD_{33}
\right]^2 \nonumber \right. \\
& &\left. +2\sumint_{Q}D_{11}\sumint_{Q}D_{11}+2\sumint_{Q}D_{22}\sumint_{Q}D_{22}+4(N-1)\sumint_{Q}D_{33}\sumint_{Q}D_{33}
\right\}\;.
\eqa
Expressing this in terms of the integrals $I_i^d$ and $I_i^T$, we obtain
\bqa
\Phi[D]&=&{\lambda\over8N}
\left\{
\left[\left(-{m_1^2\over(4\pi)^2\epsilon}+I_1^c+I_1^T\right)+\left(-{m_2^2\over(4\pi)^2\epsilon}+I_2^d+I_2^T\right)+2(N-1)\left(-{m_3^2\over(4\pi)^2\epsilon}+I_3^d+I_3^T\right)
\right]^2 \nonumber \right. \\
&&\left. +2\left(-{m_1^2\over(4\pi)^2\epsilon}+I_1^c+I_1^T\right)^2
+2\left(-{m_2^2\over(4\pi)^2\epsilon}+I_2^d+I_2^T\right)^2
+4(N-1)\left(-{m_3^2\over(4\pi)^2\epsilon}+I_3^d+I_3^T\right)^2
\right\}\;.
\label{phi01}
\eqa
The counterterm arising from ${1\over2}{\rm Tr}D_0^{-1}D$
is denoted by ${1\over2}{\rm Tr}\delta D_0^{-1}D$. Since the matrix
$\delta D_0^{-1}$ is diagonal, whose elements are given by $\delta m_i^2$,
the matrix $\delta D_0^{-1}D$ is also diagonal. We then find
\bqa
{1\over2}{\rm Tr}\delta D_0^{-1}D
&=&{1\over2}\delta m_1^2\sumint_QD_{11}
+{1\over2}\delta m_2^2\sumint_QD_{22}
+(N-1)\delta m_3^2\sumint_QD_{33}\,.
\eqa
Again the sum-integrals are known and the result is
\bqa
{1\over2}{\rm Tr}\delta D_0^{-1}D
&=&
{1\over2}\delta m_1^2\left[{m_1^2\over(4\pi)^2\epsilon}+I_1^T+I_1^c\right]
+{1\over2}\delta m_2^2\left[{m_2^2\over(4\pi)^2\epsilon}+I_2^T+I_2^c\right]
+(N-1)\delta m_3^2\left[{m_3^2\over(4\pi)^2\epsilon}+I_3^T+I_3^c\right]\;.
\label{known}
\eqa
Adding Eqs.~(\ref{zeroth}), ~(\ref{phi01}), and~(\ref{known})
and using the counterterms in Eqs~(\ref{count1})--(\ref{count3}), 
all the divergences cancel and we obtain
the renormalized effective potential to first order
\bqa\nonumber
\Omega^{(1)}&=&{1\over2}(m^2-\mu^2)\phi_0^2+{\lambda\over8N}\phi_0^4
+ \frac{1}{2}({\cal I}^c_1 + {\cal I}^T_1)
+ \frac{1}{2}({\cal I}^c_2 + {\cal I}^T_2)
+ (N-1)({\cal I}^c_3 + {\cal I}^T_3)\\ & &
\nonumber
+ {3\lambda\over 8N}({I}^c_1+I^T_1)^2
+ {3\lambda\over 8N}({I}^c_2+I^T_2)^2
+{\lambda\over 2}(N-1)({I}^c_3+I^T_3)^2 
+ {\lambda\over 4N} 
({I}^c_1+I^T_1)({I}^c_2+I^T_2)
\\  & & 
+{\lambda\over 2N}(N-1) 
({I}^c_1+I^T_1)({I}^c_3+I^T_3) 
+ {\lambda\over 2N}(N-1) 
(I^c_2+I^T_2)(I^c_3+I^T_3)\;.
\eqa
The full nonperturbative effective potential 
is now obtained by renormalizing to all
orders in the coupling constant. 
The final result is then given by Eq.~(\ref{onomega}),
where the integrals ${\cal J}_i^c$ and ${\cal J}_i^T$ are obtained from
${\cal I}_i^c$ and ${\cal I}_i^T$ 
by the replacements $m_i\rightarrow M_i$. 

\subsection{$O(2)\times O(2)$-symmetric models}

The gap equations for the dressed masses and the background field $\phi_0$
follow from 
Eqs.~(\ref{eq:D}) and~(\ref{phi2exp}):
\bqa
M^2_1 - m^2_1 &=& \delta m_0^2+{\delta\lambda^A_0 + 2 \delta\lambda^B_0\over 2}\phi_0^2
+{1\over 2}\sumint_Q\left[
3\lambda_0D_{11} + \lambda_0  D_{22} + \lambda_H D_{33}+\lambda_HD_{44}\right]
\;,\\
M^2_2-m^2_2 &=& \delta m_0^2+{\delta\lambda_0^A\over 2}\phi_0^2
+{1\over 2}\sumint_Q\left[
\lambda_0 D_{11} +  3\lambda_0D_{22} + \lambda^H D_{33}
+\lambda^HD_{44}\right]\;,\\
\label{m32x2}
M^2_3-m^2_3 &=&  \delta m_+^2+{\delta\lambda_H\over 2}\phi_0^2
+{1\over 2}\sumint_Q\left[
 \lambda_H D_{11}+\lambda_H D_{22} + 3\lambda_+D_{33} +  \lambda_+D_{44}\right]
\;,\\
\label{m42x2}
M^2_4-m^2_3 &=&  \delta m_+^2+{\delta\lambda_H\over 2}\phi_0^2
+{1\over 2}\sumint_Q\left[
 \lambda_H D_{11}+\lambda_H D_{22} + \lambda_+D_{33} +  3\lambda_+D_{44}\right]
\;,\\
0&=&\phi_0\left[\mu^2-m^2+
{3}\lambda_0\sumint_QD_{11}+\lambda_0\sumint_QD_{22}
+\lambda_H\sumint_QD_{33}+\lambda_H\sumint_QD_{44}
\right]
\;.
\eqa
Note that we have two counterterms $\lambda_0^A$  and $\lambda_0^B$ 
for $\lambda_0$ that correspond to the
two $O(2)$-invariant terms in the effective action.

Again renormalization can be carried out by the iterative procedure
of Refs.~\cite{bir3,fejos}. The first step consists of neglecting the
self-energies in the gap equations. The counterterms necessary to render
them finite are given by
\bqa
\delta m_{0(1)}^2&=&{1\over(4\pi)^2\epsilon}
\left[2\lambda_0 m_0^2 + \lambda_H m_+^2\right]\;,\\
\delta m_{+(1)}^2&=&{1\over(4\pi)^2\epsilon}
\left[2\lambda_+ m_+^2 + \lambda_H m_0^2\right]\;,\\
\delta\lambda^A_{0(1)}&=&{1\over(4\pi)^2\epsilon}
\left(3\lambda_0^2 + \lambda_H^2\right)\;,\\
\delta\lambda^B_{0(1)}&=&{\lambda_0^2\over(4\pi)^2\epsilon}\;,\\
\delta\lambda_{H(1)} &=& {2\over(4\pi)^2\epsilon}
\left(\lambda_H\lambda_0+\lambda_H\lambda_+\right)\;,
\eqa
where the index $(1)$ indicates that this is the first iteration.
The iterative procedure leads to 
the following recursion relations for the counterterms
\bqa
\delta m_{0(n)}^2&=&{1\over(4\pi)^2\epsilon}\left[
\left(\delta\lambda^A_{0(n-1)}
+\delta\lambda^B_{0(n-1)}\right) m_0^2 
+ \delta\lambda_{H(n-1)} m_+^2\right]\;,\\
\delta m_{+(n)}^2&=&{1\over(4\pi)^2\epsilon}\left[2\delta\lambda_{+(n-1)} 
m_+^2 + \delta\lambda_{H(n-1)} m_0^2\right]
\;,\\
\delta\lambda^A_{0(n)}&=&{1\over(4\pi)^2\epsilon}
\left[\lambda_0\left(2\delta\lambda^A_{0(n-1)}
+\delta\lambda^B_{0(n-1)}\right) 
+ \lambda_H\delta\lambda_{H(n-1)}\right]\;,\\
\delta\lambda^B_{0(n)}&=&
{1\over(4\pi)^2\epsilon}{\lambda_0\delta\lambda^B_{0(n-1)}}\;,\\
\delta\lambda_{+(n)}&=&{1\over(4\pi)^2\epsilon}
\left[2\lambda_+\delta\lambda_{+(n-1)}+ {1\over 2}
\lambda_H\delta\lambda_{H(n-1)}\right]\;,\\
\delta\lambda_{H(n)} &=& {1\over(4\pi)^2\epsilon}
\left[\lambda_H\left(\delta\lambda^A_{0(n-1)}
+\delta\lambda^B_{0(n-1)}\right)
+\lambda_H\left(\delta\lambda^A_{+(n-1)}
+\delta\lambda^B_{+(n-1)}\right)\right]\;.
\eqa
\end{widetext}
The dressed masses $M_3$ and $M_4$ are equal.
This follows from subtracting Eq~(\ref{m42x2}) from Eq~(\ref{m32x2}) and
renormalizing the resulting expression order by order and using the fact
that the tree-level masses $m_3$ and $m_4$ are equal.
Clearly, this holds only as long as there is no charged condensate.

These recursion relations are in agreement with those found by 
Fejos, Patkos, and Szep~\cite{fejos} 
for the $O(N)\times O(M)$-symmetry after having set
$N=M=2$. The final result for the gap equations 
is given in Eqs.~(\ref{m12pi})--(\ref{phi2pi}).
Finally, the counterterms needed to render the gap equations
finite are also those needed to renormalize the effective potential.
The final result for the effective potential is given by Eq.~(\ref{finalomega}).

\section*{Acknowledgment}
J.O.A would like to thank M. Alford and T. Brauner 
for useful discussions.


\renewcommand{\theequation}{\thesection.\arabic{equation}}

\end{document}